\documentclass[aps,floats]{revtex4}
\usepackage{amsmath,amssymb}
\usepackage{graphicx,epsfig}

\begin{document}
\bibliographystyle {plain}

\def\oppropto{\mathop{\propto}} 
\def\opsimeq{\mathop{\simeq}}
\def\opoverderline{\mathop{\overline}}
\def\operarrow{\mathop{\longrightarrow}}
\def\opsim{\mathop{\sim}}

\def\fig#1#2{\includegraphics[height=#1]{#2}}
\def\figx#1#2{\includegraphics[width=#1]{#2}}


\title{  Random Transverse Field Ising model on the Cayley Tree :  \\
analysis via Boundary Strong Disorder Renormalization  } 


 \author{ C\'ecile Monthus and Thomas Garel }
  \affiliation{Institut de Physique Th\'{e}orique, 
CNRS and CEA Saclay, 
 91191 Gif-sur-Yvette cedex, France}

\begin{abstract}

Strong Disorder Renormalization for the Random Transverse Field Ising model leads to a complicated topology of surviving clusters as soon as $d>1$. Even if one starts from a Cayley tree, the network of surviving renormalized clusters will contain loops, so that no analytical solution can be obtained. Here we introduce a modified procedure called 'Boundary Strong Disorder Renormalization' that preserves the tree structure, so that one can write simple recursions with respect to the number of generations. We first show that this modified procedure allows to recover exactly most of the critical exponents for the one-dimensional chain. After this important check, we study the RG equations for the quantum Ising model on a Cayley tree with a uniform ferromagnetic coupling $J$ and random transverse fields with support $[h_{min},h_{max}]$. We find the following picture (i) for $J>h_{max}$, only bonds are decimated, so that the whole tree is a quantum ferromagnetic cluster (ii) for $J<h_{min}$, only sites are decimated, so that no quantum ferromagnetic cluster is formed, and the ferromagnetic coupling to the boundary coincides with the partition function of a Directed Polymer model in a random medium (iii) for $h_{min}<J<h_{max}$, both sites and bonds can be decimated : the quantum ferromagnetic clusters can either remain finite (the physics is then similar to (ii), with a quantitative mapping to a modified Directed Polymer model) or an infinite quantum ferromagnetic cluster appears. We find that the quantum transition can be of two types : (a) either the quantum transition takes place in the region where quantum ferromagnetic clusters remain finite, and the singularity of the ferromagnetic coupling to the boundary involves the typical correlation length exponent $\nu_{typ}=1$ (b) or the quantum transition takes place at the point where an extensive quantum ferromagnetic cluster appears, with a correlation length exponent $\nu \simeq 0.75$.

\end{abstract}

\maketitle

\section{ Introduction }

In this paper, we consider the quantum Ising model defined in terms of Pauli matrices
\begin{eqnarray}
{\cal H} =  -  \sum_{<i,j>} J_{i,j}  \sigma^z_i \sigma^z_j - \sum_i h_i \sigma^x_i
\label{hdes}
\end{eqnarray}
where the transverse fields $h_i>0$ are independent random variables drawn with some distribution
$\pi(h_i)$ and 
where the nearest-neighbor couplings take a non-random value $J_{i,j}=J>0$
(of course, one can also consider the case where the couplings are also random,
but we have chosen to restrict to the case of non-random $J$ to simplify a bit the notations).
In dimension $d=1$, exact results for a large number of observables
have been obtained by Daniel Fisher \cite{fisher} 
 via the asymptotically exact strong disorder renormalization procedure
 (for a review, see \cite{review_strong}) : the transition is governed
by an Infinite-Disorder fixed point and 
presents unconventional scaling laws with respect to the pure case.
In dimension $d>1$, the strong disorder renormalization procedure can still be defined.
However, it cannot be solved analytically, because the topology of the lattice changes upon renormalization,
but it has been studied numerically with the conclusion that the transition is also governed by
an Infinite-Disorder fixed point in dimensions $d=2,3,4$ 
 \cite{motrunich,fisherreview,lin,karevski,lin07,yu,kovacsstrip,kovacs2d,kovacs3d,kovacsentropy,kovacsreview}
and on Erd\"os-Renyi random graphs \cite{kovacs3d,kovacsreview}.
These numerical renormalization results are in agreement with the results of independent quantum Monte-Carlo
in $d=2$ \cite{pich,rieger}.

Nevertheless, the complicated topology that emerges between renormalized degrees of freedom
in dimension $d>1$ tends to obscure the physics, because a large number of very weak bonds are generated
during the RG, that will eventually not be important for the forthcoming RG steps.
In a recent work \cite{us_boxrg}, we have thus proposed to include strong disorder RG ideas
within the more traditional fixed-length-scale real space RG framework
that preserves the topology upon renormalization : the idea is to allow some changes in the order of decimations 
with respect to the full procedure in order to maintain a simple spatial renormalized lattice. 
We have checked that in dimension $d=1$, this procedure correctly captures all critical exponents except for the 
magnetic exponent $\beta$ which is related to persistence properties of the full RG flow.
In the present paper, we introduce a similar idea for the case of the Cayley tree geometry.
If one applies the full strong disorder RG to a Cayley tree, the tree structure is rapidly destroyed
and the network of surviving clusters contains loops.
Here we thus introduce a modified procedure called Boundary Strong Disorder Renormalization
that preserves the tree structure, so that one can write simple recursions with respect to the number of 
generations.  The 'price' is again that the magnetization will not be
well taken into account, but one can hope that other scalings are correctly captured, as in \cite{us_boxrg}.
We will show that this is indeed the case in dimension $d=1$ by a direct comparison with the exact solution of the full RG procedure. For the Cayley tree, we find that when only sites are decimated within the Boundary Strong Disorder RG,
one obtains a quantitative mapping towards some Directed Polymer on the Cayley tree :
this relation with the Directed Polymer has been already obtained for the Cayley tree via some approximations
within the Quantum Cavity Approach  \cite{ioffe,feigelman,dimitrova}, and for arbitrary networks via 
simple perturbation deep in the disordered phase \cite{us_transverseDP}. However here we also consider the
possibility of bond-decimations to build quantum ferromagnetic clusters which may become important near the transition.

The paper is organized as follows.
In section \ref{sec_procedure}, we introduce the Boundary Strong Disorder Renormalization procedure
for a Cayley tree of branching ratio $K$ (i.e. each non-boundary site has $(K+1)$ neighbors).
In section \ref{sec_1d}, we show that for the one-dimensional chain corresponding to $K=1$,
this procedure is able to reproduce the most important critical exponents.
After this important check, we study the RG equation for real trees having $K>1$,
as a function of the value of the ferromagnetic $J$ with respect to the support $[h_{min},h_{max}]$
of the distribution $\pi(h_i)$ of random fields : the case $J>h_{max}$ where only bonds are decimated,
the case $J<h_{min}$ where only sites are decimated, and the case $h_{min}<J<h_{max}$ where both
bonds and sites can be decimated are discussed in sections \ref{sec_decibonds}, \ref{sec_decisites}
and \ref{sec_deciboth} respectively.

\section{ Boundary Strong Disorder Renormalization procedure }

\label{sec_procedure}

As recalled in Appendix \ref{app_full}, the Strong Disorder Renormalization
for the quantum Ising model of Eq. \ref{hdes} is {\it an energy-based RG}, 
where the strongest ferromagnetic bond or the strongest transverse field
is iteratively eliminated. In this section, we introduced a modified procedure,
called Boundary Strong Disorder RG, that preserves the tree structure,
in order to write explicit recursions for probability distributions of renormalized variables.

\subsection{ Notations }

\begin{figure}[htbp]
 \includegraphics[height=10cm]{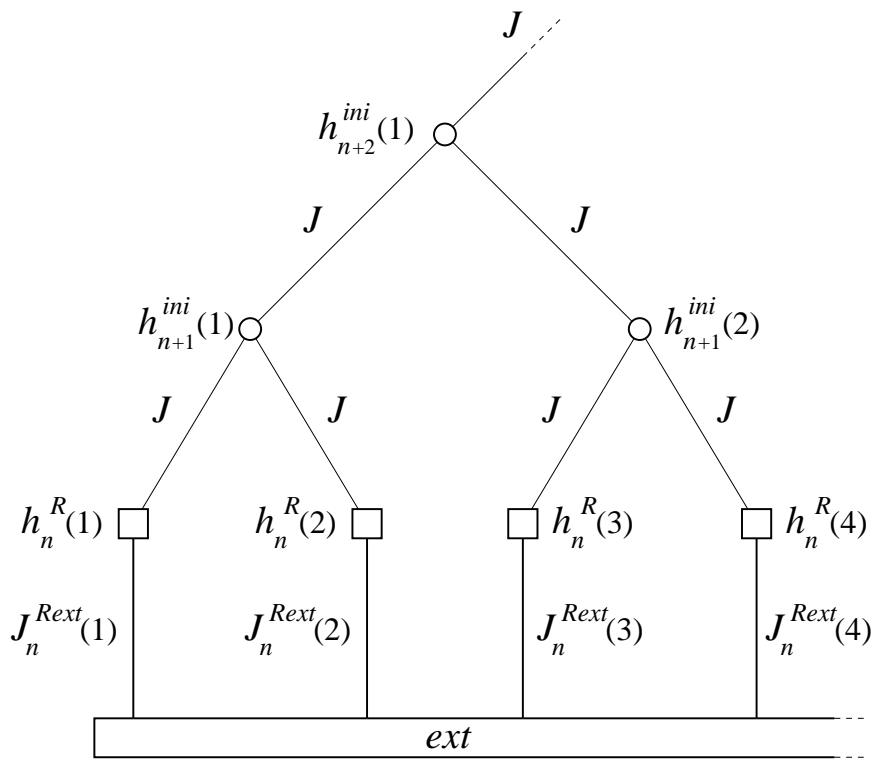}
\caption{ Renormalized structure at RG step $n$ for a Cayley tree of branching ratio $K=2$.  All sites of generations $m \geq n+1$ have not yet been modified with respect to the initial model, i.e. the sites are characterized by their initial random fields $h_m^{ini}(i)$, and are connected to their neighbors by the initial ferromagnetic coupling $J$. All sites of generations with $ m \leq n-1$ have disappeared. All sites $i$ of generation $n$ have renormalized transverse fields $h_n^{R}(i)$
and are connected to the formal external spin (called ``ext'') via some renormalized coupling $J_n^{Rext}(i)$.}
\label{figrgtree}
\end{figure}

We consider a Cayley tree of branching ratio $K$ with $N$ generations :
we label with $n=0$ the generation of external leaves that are formally connected to an exterior spin
with the coupling $J^{ext}_{(n=0)}$, in order to keep track of the coupling to the boundary upon decimation.
The next generations are labeled with $n=1,2,3,..$ that measures the distance to the boundary. 

The RG procedure consists in the renormalization of generations in the order of the label $n$
starting from $n=0$. At RG step $n$, we have thus the following structure (see Fig. \ref{figrgtree})

- all sites of generations $m \geq n+1$ have not yet been modified with respect to the initial model, i.e. the sites are characterized by their initial random fields $h_m^{ini}(i)$, and are connected to their neighbor by the initial ferromagnetic coupling $J$.

- all sites of generations with $ m \leq n-1$ have disappeared

- all sites $i$ of generation $n$ have renormalized transverse fields $h_n^{R}(i)$ that can be reduced with respect
to their initial random fields $ h^{ini}_n(i) $ by a factor $r_n(i) \leq 1$ (see Eq. \ref{hinew})
\begin{eqnarray}
h_n^{R}(i) \equiv h^{ini}_n(i) r_n(i)
\label{hrenorm}
\end{eqnarray}
and are connected to the formal external spin via some renormalized coupling $J_n^{Rext}(i)$.
These two renormalized variables  $(h^{R}_n(i), J^{Rext}_{n}(i))$ take into account 
 the formation of quantum clusters within the decimated generations $ m \leq n-1$ and the 
renormalization of couplings with the formal external spin.
However the coupling of a site $i$ of generation $n$ with its ancestor of generation $(n+1)$
has not been able to change yet and has still the value $J$ of the initial model.

To renormalize the generation $n$ that is at the boundary, we impose that all sites of generation $m \geq n+1$
and all external couplings are not decimable. Then the only variables that can be decimated
are the renormalized transverse fields $h_n^{R}(i) $ of the sites of generation $n$, and their coupling $J$
to their ancestor. Then the renormalized properties of a boundary site of generation $(n+1)$ can be computed 
from the properties of $K$ independent boundary sites of generation $n$ as we now describe.

\subsection{ Properties of a site of generation $(n+1)$ from the properties of $K$ independent sites of generation $n$ }

We consider a site of generation $(n+1)$ with an initial transverse field $h^{ini}_{n+1} $.
It is connected via the initial coupling $J$ to $K$ independent boundary sites $i=1,2..,K$ of generation $n$,
that have renormalized transverse fields  $h_n^{R}(i)= h^{ini}_n(i) r_n(i)$
and renormalized external couplings $J^{Rext}_{n}(i)$

Since the site of generation $(n+1)$ is declared to be undecimable for the moment,
we may first renormalize independently each branch $i=1,2..,K$

\subsubsection { RG rule for each branch $i=1,2,..,K$ }

\label{sec_rgbranch}

For each branch $i=1,2,..K$, we apply the Strong Disorder RG rules as follows (see Appendix \ref{app_full}).
We consider the maximum between the renormalized transverse field $h^{R}_n(i)=h^{ini}_n(i) r_n(i)$ and the coupling $J$
to the site of generation $(n+1)$ 

(a) if $ h^{R}_n(i)= h^{ini}_n(i) r_n(i) > J$ : we decimate the site $i$ (see case (i) in Appendix \ref{app_full})
so that the site of generation $(n+1)$ will get no contribution from branch $i$ to renormalize its transverse field
\begin{eqnarray}
r_n^{new}(i) && = 1
\label{hmua}
\end{eqnarray}
whereas the contribution of branch $i$ to the external coupling will be (see Eq A2)
\begin{eqnarray}
J_n^{new}(i)= J^{Rext}_n(i) \frac{ J }{h^{R}_n(i) } = J^{Rext}_n(i) \frac{ J }{h^{ini}_n(i) r_n(i) }
\label{jextnewa}
\end{eqnarray}

(b) if $ h^{R}_n(i)\equiv h^{ini}_n(i) r_n(i) < J$ : we decimate the corresponding bond (see case (ii) in Appendix \ref{app_full}), i.e. the site $i$ of generation $n$ is merged with its ancestor of generation $(n+1)$.
The contribution of site $i$ to the renormalization of the transverse field of generation $(n+1)$
corresponds to the reducing factor (see Eq. A3)
\begin{eqnarray}
r_n^{new}(i) =    \frac{h^{R}_n(i) }{ J } = r_n(i) \frac{ h^{ini}_n(i) }{ J }
\label{hmub}
\end{eqnarray}
whereas the contribution of branch $i$ to the external coupling will be 
\begin{eqnarray}
J_n^{new}(i)= J^{Rext}_n(i)
\label{jextnewb}
\end{eqnarray}

\subsubsection { Taking into account the global effect of the RG of the $K$ branches }

We now have to take into account the global effect of the $K$ independent branches 
to obtain the renormalized properties of the site of generation $(n+1)$
\begin{eqnarray}
r_{n+1} && = \prod_{i=1}^K r_n^{new}(i) 
\nonumber \\
J^{Rext}_{n+1} && = \sum_{i=1}^K J_n^{new}(i)
\label{bilan2branches}
\end{eqnarray}

\subsection { RG equation for the joint probability distribution $P_n(J^{ext},r)$ }

The above rules may be summarized by the following equation for the joint probability distribution
$P_n(J^{ext},r)$ of the renormalized external coupling $J^{ext}=J^{Rext}_n(i)$ 
and the reducing factor $r=h^R_n(i)/h^{ini}_n(i) $ of the renormalized transverse field for a site $i$
of the boundary generation $n$
\begin{eqnarray}
P_{n+1}(J^{ext},r) && = \prod_{i=1}^K \int d J^{ext}_i dr_i P_{n}(J^{ext}_i,r_i) \int dh^{ini}_i \pi(h^{ini}_i )
\int d J^{new}_i d r^{new}_i 
\nonumber \\
&& \prod_{i=1}^K \left[ \theta( h^{ini}_i r_i \geq J) 
\delta(r^{new}_i -1  ) \delta(J^{new}_i - J^{ext}_i \frac{J}{h^{ini}_i r_i } )
+ \theta( h^{ini}_i r_i <J) \delta(r^{new}_i - r_i \frac{h^{ini}_i}{J}  ) \delta(J^{new}_i - J^{ext}_i  ) \right]
\nonumber \\
&& \delta \left(J^{ext}- \sum_{i=1}^K J^{new}_i \right)
\delta \left (r-  \prod_{i=1}^K  r^{new}_i  \right)
\label{evolPK}
\end{eqnarray}
The first line corresponds to the drawing of $K$ independent sites of generation $n$ with their renormalized variables
of joint probability  $P_{n}(J^{ext}_i,r_i) $ and to the drawing of $K$ initial transverse fields $h^{ini}_i $ 
with the law $\pi(h^{ini}_i)$.
The second lign takes into account the renormalization within each branch $(i)$ 
that give the contribution  $(J^{new}_i , r^{new}_i )$.
Finally the third line takes into account the global effect of the $K$ branches.
The initial condition is simply
\begin{eqnarray}
P_{n=0}(J^{ext},r) && = \delta(J^{ext}-J^{ext}_0) \delta(r-1)
\label{PKinitial}
\end{eqnarray}

For a finite Cayley Tree containing $N$ generations, the final state after $(N-1)$ RG steps corresponds
to the single Center site which will have a renormalized transverse field given by a reducing factor $r_{center}$
and a renormalized coupling $J^{ext}_{center}$ given by taking into account $(K+1)$ independent branches
of law $P_{N-1}(J^{ext},r)$ obtained from Eq. \ref{evolPK}
\begin{eqnarray}
&& P_{N}^{center}(J^{ext}_{center},r_{center})  = \prod_{i=1}^{K+1} \int d J^{ext}_i dr_i P_{n}(J^{ext}_i,r_i) \int dh^{ini}_i \pi(h^{ini}_i )
\int d J^{new}_i d r^{new}_i 
\nonumber \\
&& \prod_{i=1}^{K+1} \left[ \theta( h^{ini}_i r_i \geq J) 
\delta(r^{new}_i -1  ) \delta(J^{new}_i - J^{ext}_i \frac{J}{h^{ini}_i r_i } )
+ \theta( h^{ini}_i r_i <J) \delta(r^{new}_i - r_i \frac{h^{ini}_i}{J}  ) \delta(J^{new}_i - J^{ext}_i  ) \right]
\nonumber \\
&& \delta \left(J^{ext}_{center}- \sum_{i=1}^{K+1} J^{new}_i \right)
\delta \left (r_{center}-  \prod_{i=1}^{K+1}  r^{new}_i  \right)
\label{center}
\end{eqnarray}

\subsection { Pure model without disorder }

\label{sec_pure}

To get some idea of the meaning of the RG Eq. \ref{evolPK},
let us first consider the {\it pure } quantum Ising model
 with uniform transverse field $h_0$ and uniform coupling $J$.

The boundary RG procedure described above gives the following results :

(a) for $h_0<J$, only bonds are decimated, and the whole tree is a single ferromagnetic clusters which is connected to the external via the renormalized coupling that grows exponentially in $n$
\begin{eqnarray}
J_n^{ext}=  K^n J_{0}^{ext}
\label{jextnewbpur}
\end{eqnarray}

(b) for $h_0>J$, only sites are decimated so that no quantum clusters are formed, and the transverse fields are not renormalized. The external coupling after $n$ generations behaves as
\begin{eqnarray}
J_{n}^{ext} && =  J_{n-1}^{ext} \left( K  \frac{ J }{h_0 } \right)
= J_{0}^{ext} \left( K  \frac{ J }{h_0 } \right)^n
\label{bilanpurdes}
\end{eqnarray}
So even if each branch yields a reducing factor $J/h_0<1$,  the branching ratio $K$ can overcompensate
this one-dimensional decay.
In this analysis, the pure transition thus takes place at
\begin{eqnarray}
J_{c}^{pure} && = \frac{h_0 }{K} 
\label{jcpure}
\end{eqnarray}
which coincides with the transition found within the 'Cavity Mean-Field' approach 
(see the discussion on various approximation in \cite{dimitrova}).

The disordered phase $h_0>K J_0$ then correspond to the exponential decay
\begin{eqnarray}
J_n^{ext}= J_{0}^{ext} e^{- \frac{n}{\xi_{pure}}} \ \ { \rm with } \ \ \xi_{pure} = \frac{1}{\ln \frac{J_c}{J}} \propto (J_c-J)^{-\nu_{pure}}
\label{jcpuredecay}
\end{eqnarray}
with the correlation length exponent
\begin{eqnarray}
\nu_{pure}=1
\label{nujcpuredecay}
\end{eqnarray}
The ordered phase $h_0<K J_0$ correspond to the exponential growth
 $J_n^{ext}= J_{0}^{ext} \left(   \frac{ J }{ J_c } \right)^n$.
We should stress here that the diverging correlation length $\xi_{pure} $ defined from $J_n^{ext}$
should not be confused with the correlation length $\xi_{2} $ governing the decay of the two-point correlation function
between two points at distance $n$
\begin{eqnarray}
 C(n) \propto \left(   \frac{ J }{h_0 } \right)^n = e^{- \frac{n}{\xi_2}}
\label{crpure}
\end{eqnarray}
The comparison with Eq. \ref{bilanpurdes} shows that the two correlation lengths are related via
\begin{eqnarray}
\frac{1}{\xi_{pure}} = \frac{1}{\xi_{2}} - \ln K
\label{pur2xi}
\end{eqnarray}
coming from the exponential number $K^n$ of points at generation $n$.
In particular at criticality when $\xi_{pure}$ diverges $\xi_{pure} \to +\infty $, the  correlation length $\xi_{2} $
takes the finite value $\xi_{2}^{criti} = \frac{1}{\ln K} $ in agreement with the studies \cite{nagaj,nagy}.

Note that for the pure case, the present method based
 on second order perturbation theory in 
 $J/h_0$ will be more and more justified in the critical region in the limit of large
branching ratio $K$ since the transition takes place at smaller and smaller value of the ratio $J_c/h_0 \sim 1/K \to 0$.

\subsection{ Relevance of small disorder around the pure transition }

\label{sec_harris}

In finite dimension below the upper critical dimension $d<d_u$,
the relevance of small disorder at the transition of the pure model
is determined by the Harris criterion \cite{harris}  
or the inequality $\nu \geq 2/d$ of Chayes {\it et al} \cite{chayes}.
Above the upper-critical dimension $d \geq d_u$ of the pure model,
the Harris criterion is more involved \cite{nishiyama,sarlat,us_transverseDP}
since there is no hyperscaling
and there are two distinct correlation length exponents. 
For the special case of the Cayley tree that corresponds formally to $d \to +\infty$,
the naive adaptation of the Harris criterion yields that small disorder is always
irrelevant at the pure transition for the following reason : 
the correlated volume $\xi^d$ in finite dimension $d$ becomes $V=K^{\xi}$ on the Cayley
tree of branching ratio $K$; so the Central-Limit disorder fluctuations which are 
of order $1/\sqrt{\xi^d}$ in finite dimension $d$ becomes exponentially small
of order $1/\sqrt{K^{\xi}}$ on
the Cayley tree. So these disorder fluctuations 
are always much smaller than the power-law distance to criticality
of the pure model $(J_c-J) \propto \xi^{-1/\nu_{pure}} $ (Eq. \ref{jcpuredecay}).

Note however that for quantum disordered models, 
these arguments based on the weak-disorder Harris criterion have been questioned \cite{motrunich,fisherreview} since rare regions can play an essential role since the disorder is
actually 'infinitely' correlated along the time-direction. The effects of rare regions are discussed in detail in the
review \cite{vojta}, with the conclusion that the important parameter is the effective dimensionality $d_{RR}$
of rare regions (in our present quantum model where the disorder is actually 'infinitely' 
correlated along the time-direction, the dimensionality of rare regions is $d_{RR}=1$ \cite{vojta}).
This dimensionality $d_{RR}$ of rare regions should be compared to the lower critical dimension $d^{class}_l$
sufficient to obtain ordering  (in our present quantum model where the order is a ferromagnetic magnetization,
the lower critical dimension is $d^{class}_l=1$). The random quantum Ising model thus corresponds to 
the case $d_{RR}=d^{class}_l$, i.e. to the so-called class B in the classification described in \cite{vojta},
 where rare regions dominate the critical behavior and can 
induce an unusual activated scaling.
(The conventional power-law scaling is expected
 to hold for the so-called class A corresponding to $d_{RR}<d^{class}_l$, 
where rare regions cannot undergo the phase transition by themselves, 
whereas the so-called class C corresponds to the case $d_{RR}>d^{class}_l$
 where rare regions can order by themselves
at different values of the order parameter).

\subsection{ Qualitative description of the possible scenarios for the random critical point }

Before entering the technical solutions of the RG Equation \ref{evolPK}
in various phases and regimes, let us now briefly summarize 
the two possible scenarios that we will find
for the transition of the disordered model on the Cayley
tree of branching ratio $K>1$ :

(a) either the random quantum transition takes place in the region where quantum ferromagnetic clusters do not exist (as in the pure case described in section \ref{sec_pure} above) or remain finite.
Then the ferromagnetic coupling to the boundary will be found to
behave in the disordered phase as 
$J^{ext}_n \sim e^{- n/\xi_{typ}}$
where the correlation length $\xi_{typ}$ diverges with the typical correlation 
length exponent $\nu_{typ}=1$ which coincides with the pure exponent of Eq. \ref{nujcpuredecay}. This case can be thus understood as a 'weak-disorder' case where the critical exponents of the pure transition are not changed by the disorder, in agreement with the naive
adaptation of the Harris criterion discussed in section \ref{sec_harris} above.

(b) or the quantum transition takes place at the point 
where an extensive quantum ferromagnetic cluster appears, and is thus completely different
from the pure transition on the tree. We will obtain below that the
renormalized transverse field scales as $\ln h_R(n) \propto  - K^{n-n_*}$ 
for $n$ generations, where the distance $n_*$ diverges with a correlation length exponent
of order $\nu_* \simeq 0.75$.

\section { Boundary Strong Disorder RG for the one-dimensional chain $K=1$ }

\label{sec_1d}

Before studying the case of real trees having $K>1$, we consider in the present section
the case of the one-dimensional chain corresponding to $K=1$ to check the validity
of the Boundary Strong Disorder RG procedure by comparison with the exact solution
of the usual full RG equation \cite{fisher}.

\subsection{ Choice of Binary disorder }

To simplify the technical details, we now focus on the binary distribution
for the initial transverse fields
\begin{eqnarray}
\pi(h^{ini} ) = p_1 \delta(h^{ini}-h_1) + p_2 \delta(h^{ini}-h_2) \ \ {\rm with } \ \ p_1+p_2=1
\label{binary}
\end{eqnarray}
and we also fix the ferromagnetic coupling to the value $J=\sqrt{h_1 h_2} \in ]h_1,h_2[$ 
i.e. 
\begin{eqnarray}
 \frac{h_1}{J} = \frac{J}{h_2} = \sqrt{\frac{h_1}{h_2}} <1
\label{slopea1}
\end{eqnarray}
so that the control parameter of the transition is the fraction $p_2=1-p_1$ of Eq. \ref{binary}.
The exact criterion for the critical coupling $J_c$ \cite{pfeuty}
\begin{eqnarray}
\ln J_c= \overline{\ln h} = p_1 \ln h_1 + p_2 \ln h_2
\label{exactcriti1d}
\end{eqnarray}
leads here with Eq. \ref{slopea1} to the critical point
\begin{eqnarray}
p_1^{criti}=1-p_2^{criti}=\frac{1}{2}
\label{exactcriti1dbis}
\end{eqnarray}

Starting from the initial condition of Eq \ref{PKinitial},
it is easy to see that the only renormalized values that can appear are 
the discrete values
\begin{eqnarray}
 r_{k} \equiv \left( \frac{h_1}{J}  \right)^{k} \ {\rm with } \ \ k=0,1,2,..
\label{rk}
\end{eqnarray}
and 
\begin{eqnarray}
 J^{ext}_m \equiv  J^{ext}_0 \left( \frac{h_1}{J}  \right)^{m} \ {\rm with } \ \ m=0,1,2,..
\label{jm}
\end{eqnarray}

i.e. the joint probability distribution $P_n(J^{ext},r)$ takes the form of a 
double sum of delta functions
\begin{eqnarray}
 P_n(J^{ext},r)= \sum_{m=0}^{+\infty} \sum_{k=0}^{+\infty} c_n(m,k) 
 \delta \left[J^{ext} - J^{ext}_0 \left( \frac{h_1}{J}  \right)^{m}  \right] \delta \left[ r- \left( \frac{h_1}{J}  \right)^{k}  \right]
\label{ansatmk}
\end{eqnarray}
with the normalization 
\begin{eqnarray}
1=  \int dJ^{ext} \int_0^1 dr  P_n(J^{ext},r) = \sum_{m=0}^{+\infty} \sum_{k=0}^{+\infty} c_n(m,k) 
\label{normacmk}
\end{eqnarray}
and the initial condition (Eq \ref{PKinitial} )
\begin{eqnarray}
 c_{n=0}(m,k) =\delta_{m,0} \delta_{k,0}
\label{initialcmk}
\end{eqnarray}

Plugging the form of Eq. \ref{ansatmk} into the RG Equation of Eq. \ref{evolPK}
yields that the double generation function
\begin{eqnarray}
{\hat c}_n(y,z) \equiv \sum_{m=0}^{+\infty} \sum_{k=0}^{+\infty} y^m z^k c_n(m,k) 
\label{defcyz}
\end{eqnarray}
satisfies the linear recurrence
\begin{eqnarray}
{\hat c}_{n+1}(y,z) =  \left[  p_1  z +  \frac{p_2}{z}  \right]{\hat c}_{n}(y,z)
+ p_2 \left[ y  -  \frac{1}{z}   \right] {\hat c}_{n} (y,0) 
\label{reccyz}
\end{eqnarray}
with the initial condition (Eq \ref{initialcmk})
\begin{eqnarray}
  {\hat c}_{n=0} (y,z) = 1
\label{hatCiniyz}
\end{eqnarray}

\subsection{ Solution for arbitrary finite-size $n$ }

Let us introduce the generating function with respect to the size $n$

\begin{eqnarray}
G(y,z,\lambda) \equiv \sum_{n=0}^{+\infty} \lambda^n  {\hat c}_{n} (y,z) 
\label{defG}
\end{eqnarray}
Summing Eq \ref{reccyz} over $n$ after multiplying by $\lambda^n$,
yields with the initial condition of Eq \ref{hatCiniyz}
\begin{eqnarray}
\sum_{n=0}^{+\infty} \lambda^n  {\hat c}_{n+1} (y,z)&&  
= \frac{1}{\lambda} \left[ \sum_{m=1}^{+\infty} \lambda^m  {\hat c}_{m} (y,z) \right] 
=  \frac{1}{\lambda} \left[ G(y,z,\lambda) - 1 \right]
 \nonumber \\ 
&& = G(y,z,\lambda) \left[  p_1 z +  \frac{p_2}{z}  \right]
+ p_2   G (y,0,\lambda) \left[ y  -  \frac{1}{z}   \right]
\label{recgenec1dC}
\end{eqnarray}
i.e. $G(y,z,\lambda) $ satisfies the linear equation
\begin{eqnarray}
\left[ 1 -\lambda ( p_1  z +  \frac{p_2}{z})  \right]   G(y,z,\lambda)
+ \lambda p_2 G (y,0,\lambda) \left[   \frac{1}{z}  - y   \right]
 - 1  
 = 0
\label{eqlinearG1d}
\end{eqnarray}
of kernel
\begin{eqnarray}
 N(z,\lambda) =  1 -\lambda ( p_1  z +  \frac{p_2}{z})  
\label{noyau}
\end{eqnarray}
The two roots of this kernel reads
\begin{eqnarray}
 Z^{\pm}(\lambda) = \frac{1 \pm \sqrt{1 - 4 \lambda^2 p_1 p_2  } }{2 \lambda p_1  }  
\label{solunoyau}
\end{eqnarray}
The solution which is regular for $\lambda \to 0$ is $Z^{-}(\lambda)  $.
Replacing $z$ by $Z^{-}(\lambda) $ in Eq. \ref{eqlinearG1d}
yields
\begin{eqnarray}
 \lambda p_2  G (y ,0,\lambda) \left[   \frac{1}{Z^{-}(\lambda)} - y   \right]
 - 1  
 = 0
\label{eqlinearGsolunoyauz}
\end{eqnarray}
i.e.
\begin{eqnarray}
  G (y,0,\lambda) = \frac{Z^{-}(\lambda)}{ \lambda p_2 (1 - y  Z^{-}(\lambda))}
\label{eqlinearGsolunoyau}
\end{eqnarray}

We may now plug this into Eq \ref{eqlinearG1d} to obtain the full solution
\begin{eqnarray}
   G(y,z,\lambda)   
&&  = \frac{ 1 - \lambda p_2 G (y,0,\lambda) \left[   \frac{1- y z}{z}     \right]}{\left[ 1 -\lambda ( p_1  z + p_2 \frac{1}{z})  \right]}
 = \frac{  \frac{(z- Z^{-}(\lambda))  }{  (1- y Z^{-}(\lambda)) z } }
{ \frac{(- \lambda p_1)}{z} \left[z- Z^{-}(\lambda)\right]  \left[z- Z^{+}(\lambda) \right]}
 = \frac{ 1 }
{ ( \lambda p_1) \left[ (1- y Z^{-}(\lambda))   ( Z^{+}(\lambda))- z \right]}
\label{eqlinearG1dsol}
\end{eqnarray}
The expansion in $y$ and $z$ yields the generating function of the joint distribution $c_n(m,k)$
\begin{eqnarray}
{\tilde c}_{\lambda}(m,k) \equiv \sum_{n=0}^{+\infty} \lambda^n  c_n(m,k)
= \frac{ 1 }
{  \lambda p_1  }
\left[ Z^{-}(\lambda) \right]^m \left( \frac{1}{Z^{+}(\lambda)} \right)^{k+1}
\label{tilde}
\end{eqnarray}

The joint distribution $c_n(m,k) $ for any finite-size $n$ can be thus obtained by inversion
\begin{eqnarray}
c_n(m,k) && = \oint_{C_0} \frac{d \lambda }{2i \pi \lambda^{n+1} } {\tilde c}_{\lambda}(m,k)
= \oint_{C_0} \frac{d \lambda }{2i \pi \lambda^{n+1} }
\frac{ 1 }
{  \lambda p_1  }
\left[ \frac{1 - \sqrt{1 - 4 \lambda^2 p_1 p_2  } }{2 \lambda p_1  }  \right]^m 
\left[  \frac{2 \lambda p_1  }{1 + \sqrt{1 - 4 \lambda^2 p_1 p_2  } } \right]^{k+1}
 \nonumber \\ && 
=  \oint_{C_0} \frac{d \lambda }{2i \pi \lambda^{n+1} }
\frac{ 1 }
{  \lambda p_1  }
\left[ \frac{1 - \sqrt{1 - 4 \lambda^2 p_1 p_2  } }{2 \lambda p_1  }  \right]^m 
\left[  \frac{1 - \sqrt{1 - 4 \lambda^2 p_1 p_2  } }{2 \lambda p_2  }\right]^{k+1}
 \nonumber \\ && 
= \frac{ 1 }
{   p_1^{1+m} p_2^{k+1}  } \oint_{C_0} \frac{d \lambda }{2i \pi \lambda^{n+(m+1)+(k+1)} }
\left[ \frac{1 - \sqrt{1 - 4 \lambda^2 p_1 p_2  } }{2   }  \right]^{m+k+1}
\label{invGzerozero}
\end{eqnarray}
where
 $C_0$ is a circle around the origin $\lambda=0$ in the complex plane.

\subsection{ Asymptotic distribution of renormalized transverse fields as $n \to +\infty$ }

Let us now consider the probability distribution $P_n(r)$ of the variable $r$ alone
\begin{eqnarray}
P_n(r) \equiv \int dJ^{ext} P_n(J^{ext},r)
\label{ralone}
\end{eqnarray}
In terms of the representation of Eq. \ref{ansatmk},
this corresponds to the probability distribution of the integer $k$ alone
\begin{eqnarray}
c_n(k) \equiv \sum_{m=0}^{+\infty}  c_n(m,k) 
\label{kalone}
\end{eqnarray}
From Eq. \ref{defG} and \ref{eqlinearG1dsol}, its generating function reads
\begin{eqnarray}
\sum_{n=0}^{+\infty} \lambda^n \sum_{k=0}^{+\infty} z^k c_n(k) = 
G(y=1,z,\lambda) 
 = \frac{ 1 }
{ ( \lambda p_1) \left[1-  Z^{-}(\lambda)\right]   \left[ Z^{+}(\lambda))- z \right]}
\label{genekalone}
\end{eqnarray}
The series expansion in $z$ yields
\begin{eqnarray}
\sum_{n=0}^{+\infty} \lambda^n  c_n(k)  
&&  = \frac{ 1 }
{ ( \lambda p_1) \left[1-  Z^{-}(\lambda)\right]   \left[ Z^{+}(\lambda)) \right]^{k+1}}
 \nonumber \\
 && = 
 \frac{ 2 }
{  \left[ 2 \lambda p_1 - 1 + \sqrt{1 - 4 \lambda^2 p_1 p_2  }   \right]}
 \left[ \frac{1 - \sqrt{1 - 4 \lambda^2 p_1 p_2  } }{2 \lambda p_2  }
  \right]^{k+1} 
\label{genekalonek}
\end{eqnarray}
so that after inversion, one obtains
\begin{eqnarray}
c_n(k) 
 && = \oint_{C_0} \frac{d \lambda }{ i \pi \lambda^{n+1} }
 \frac{ 1 }
{  \left[ 2 \lambda p_1 - 1 + \sqrt{1 - 4 \lambda^2 p_1 p_2  }   \right]}
 \left[ \frac{1 - \sqrt{1 - 4 \lambda^2 p_1 p_2  } }{2 \lambda p_2  }
  \right]^{k+1} 
\label{kaloneres}
\end{eqnarray}
The complex function contains two cuts $]-\infty,-\lambda_c]$ and $[\lambda_c,+\infty[$
on the real axis with
\begin{eqnarray}
\lambda_c = \frac{1}{2 \sqrt{p_1 p_2} } = \frac{1}{2 \sqrt{(1-p_2) p_2} } \geq 1
\label{lambdac}
\end{eqnarray}
In addition, there exists a simple pole at
\begin{eqnarray}
\lambda_P=1 \ \ \ \ \ \ {\rm if } \ \ p_1=1-p_2 < \frac{1}{2}
\label{solupoleGcarre}
\end{eqnarray}
In this region $p_1=1-p_2 < \frac{1}{2}$, one thus obtains after deformation of the contour in 
the complex plane that $c_n(k) $ converges towards a finite limit $ c_{\infty}(k)$
given by the residue of the pole at $ \lambda_P=1$ yielding
\begin{eqnarray}
c^{disordered}_n(k) && \oppropto_{n \to +\infty}  c^{disordered}_{\infty} (k) 
= \frac{1-2 p_1}{1-p_1} \left(\frac{p_1}{1-p_1}\right)^k  \ \ \ \ {\rm if } \ \ p_1 < \frac{1}{2}
\label{solucnkfinite1d}
\end{eqnarray}

In the complementary region $p_1=1-p_2 > \frac{1}{2}$, the variable $k$ does not remain finite as $n \to +\infty$, and its asymptotic behavior will be governed by the singularities near the branching points $(\pm \lambda_c)$ of the two cuts of Eq. \ref{kaloneres}. 
However, to see more directly the appropriate scaling limit, it is more convenient
to treat the large integers $n$ and $k$ as real variables, and to replace finite sums by integrals,
so that the generating function in $\lambda<1$ becomes a Laplace transform in the variable
$s= - \ln \lambda \geq 0$ (Eq \ref{genekalonek})
\begin{eqnarray}
\int_0^{+\infty} dn e^{-s n} c_n(k) && \simeq 
\left[ \sum_{n=0}^{+\infty} \lambda^n  c_n(k) \right]_{\lambda=e^{-s}} 
   \nonumber \\
 && \simeq
 \frac{ 2 }
{  \left[ 2  p_1 e^{-s}  - 1 + \sqrt{1 - 4  p_1 p_2 e^{-2 s} }   \right]}
 \left[ \frac{1 - \sqrt{1 - 4 p_1 p_2 e^{-2s} } }{2 p_2 e^{-s} }
  \right]^{k+1} 
\label{genekaloneklaplace}
\end{eqnarray}
To study the asymptotic behavior of $c_n(k)$ as $n \to +\infty$,
we have to determine the leading behavior as $s \to 0$ of the right-handside.

Let us first consider the critical case $p_1^{criti}=p_2^{criti}=\frac{1}{2}$, where Eq \ref{genekaloneklaplace}
becomes
\begin{eqnarray}
\int_0^{+\infty} dn e^{-s n} c^{criti}_n(k) 
 && \simeq
 \frac{ 2 }
{  \left[  e^{-s}  - 1 + \sqrt{1 -  e^{-2 s} }   \right]}
 \left[ \frac{1 - \sqrt{1 -  e^{-2s} } }{ e^{-s} } \right]^{k+1} 
  \nonumber \\
 &&
\oppropto_{s \to 0} \sqrt{\frac{2}{s}} e^{ -k \sqrt{2s}}
\label{genekalonekcriti}
\end{eqnarray}
i.e. after inversion of this Laplace Transform, one obtains that $c_n(k)$ is a half Gaussian
defined for $k>0$
\begin{eqnarray}
 c^{criti}_n(k) 
\oppropto_{n \to +\infty} \theta (k \geq 0) \sqrt{\frac{2}{\pi n}} e^{ - \frac{k^2}{2n}}
\label{cnkcriti}
\end{eqnarray}

Let us now consider the ordered phase $p_1=1-p_2>\frac{1}{2}$, where Eq \ref{genekaloneklaplace}
becomes in the limit $s \to 0$
\begin{eqnarray}
\int_0^{+\infty} dn e^{-s n} c^{ordered}_n(k) 
 && 
\oppropto_{s \to 0} \frac{1}{(2 p_1-1)} 
e^{ k \left[ - \frac{s}{(2 p_1-1)} + \frac{2 p_1 (1-p_1) s^2}{(2 p_1-1)^3}\right]}
\label{cnkoffcritigene}
\end{eqnarray}
i.e. after Laplace inversion, 
one obtains that $c_n(k)$ is asymptotically a Gaussian distribution
\begin{eqnarray}
 c_n^{ordered}(k) 
\oppropto_{n \to +\infty} \frac{1}{ \sqrt{8 \pi p_1 p_2  n}} e^{ - \frac{[k-(2 p_1-1)n ]^2}{ 8 p_1 p_2 n}}
\label{cnkoffcriti}
\end{eqnarray}

\subsection{ Asymptotic distribution of renormalized couplings $J^{ext}$ as $n \to +\infty$ }

Let us now consider the probability distribution $Q_n(J^{ext})$ of the variable $J^{ext}$ alone
\begin{eqnarray}
Q_n(J^{ext}) \equiv \int dr P_n(J^{ext},r)
\label{raloneQ}
\end{eqnarray}
In terms of the representation of Eq. \ref{ansatmk},
this corresponds to the probability distribution of the integer $m$ alone
\begin{eqnarray}
q_n(m) \equiv \sum_{k=0}^{+\infty}  q_n(m,k) 
\label{kaloneq}
\end{eqnarray}
From Eq. \ref{defG} and \ref{eqlinearG1dsol}, its generating function reads
\begin{eqnarray}
\sum_{n=0}^{+\infty} \lambda^n \sum_{m=0}^{+\infty} y^m q_n(m) = 
G(y,z=1,\lambda) 
 = \frac{ 1 }
{ ( \lambda p_1) \left[1- y Z^{-}(\lambda)\right]   \left[ Z^{+}(\lambda))- 1 \right]}
\label{genemalone}
\end{eqnarray}
The series expansion in $y$ yields
\begin{eqnarray}
\sum_{n=0}^{+\infty} \lambda^n  q_n(m)  
&&  = \frac{ \left[ Z^{-}(\lambda)\right]^m }
{ ( \lambda p_1)    \left[ Z^{+}(\lambda))-1 \right]}
  = \frac{ 2 }
{     \left[1 + \sqrt{1 - 4 \lambda^2 p_1 p_2  }- 2 \lambda p_1   \right]}\left[ \frac{1 - \sqrt{1 - 4 \lambda^2 p_1 p_2  } }{2 \lambda p_1  } \right]^m
\label{genemalonem}
\end{eqnarray}

so that after inversion, one obtains
\begin{eqnarray}
q_n(m) 
 && = \oint_{C_0} \frac{d \lambda }{ i \pi \lambda^{n+1} }
\frac{ 1 }
{     \left[1 + \sqrt{1 - 4 \lambda^2 p_1 p_2  }- 2 \lambda p_1   \right]}\left[ \frac{1 - \sqrt{1 - 4 \lambda^2 p_1 p_2  } }{2 \lambda p_1  } \right]^m
\label{maloneres}
\end{eqnarray}
The complex function contains two cuts $]-\infty,-\lambda_c]$ and $[\lambda_c,+\infty[$
on the real axis with Eq \ref{lambdac}.
In addition, there exists a simple pole at
\begin{eqnarray}
\lambda_P=1 \ \ \ \ \ \ {\rm if } \ \ p_1=1-p_2 > \frac{1}{2}
\label{solupoleGcarrebis}
\end{eqnarray}
In this region $p_1=1-p_2 > \frac{1}{2}$, one thus obtains after deformation of the contour in 
the complex plane that $q_n(m) $ converges towards a finite limit $ q_{\infty}(m)$
given by the residue of the pole at $ \lambda_P=1$ yielding
\begin{eqnarray}
q^{ordered}_n(m) && \oppropto_{n \to +\infty}  q^{ordered}_{\infty} (m) 
= \frac{2 p_1-1}{p_1} \left(\frac{1-p_1}{p_1}\right)^m  \ \ \ \ {\rm if } \ \ p_1 > \frac{1}{2}
\label{solucnkfinite1dm}
\end{eqnarray}

In the complementary region $p_1=1-p_2 < \frac{1}{2}$, the variable $m$ does not remain finite as $n \to +\infty$, and its asymptotic behavior will be governed by the singularities near the branching points $(\pm \lambda_c)$ of the two cuts of Eq. \ref{maloneres}. 
However, to see more directly the appropriate scaling limit, it is more convenient
to treat the large integers $n$ and $m$ as real variables, and to replace finite sums by integrals,
so that the generating function in $\lambda<1$ becomes a Laplace transform in the variable
$s= - \ln \lambda \geq 0$ (Eq \ref{genemalonem})
\begin{eqnarray}
\int_0^{+\infty} dn e^{-s n} q_n(m) && \simeq 
\left[ \sum_{n=0}^{+\infty} \lambda^n  q_n(m) \right]_{\lambda=e^{-s}} 
   \nonumber \\
 && \simeq
\frac{ 2 }
{     \left[1 + \sqrt{1 - 4  p_1 p_2 e^{-2s} }- 2  p_1 e^{-s}  \right]}
\left[ \frac{1 - \sqrt{1 - 4  p_1 p_2 e^{-2 s} } }{2  p_1 e^{-s} } \right]^m
\label{laplacem}
\end{eqnarray}
To study the asymptotic behavior of $q_n(m)$ as $n \to +\infty$,
we have to determine the leading behavior as $s \to 0$ of the right-handside.

Let us first consider the critical case $p_1=p_2=\frac{1}{2}$, where Eq \ref{laplacem}
becomes
\begin{eqnarray}
\int_0^{+\infty} dn e^{-s n} q^{criti}_n(m) 
 && \simeq  \frac{ 2 }
{     \left[1 + \sqrt{1 -  e^{-2s} }-  e^{-s}  \right]}
\left[ \frac{1 - \sqrt{1 -  e^{-2 s} } }{ e^{-s} } \right]^m
  \nonumber \\
 &&
\oppropto_{s \to 0} \sqrt{\frac{2}{s}} e^{ - m \sqrt{2s}}
\label{genemalonekcriti}
\end{eqnarray}
i.e. after inversion of this Laplace Transform, one obtains that $q_n(m)$ is a half Gaussian
defined for $k>0$
\begin{eqnarray}
 q^{criti}_n(m) 
\oppropto_{n \to +\infty} \theta (m \geq 0) \sqrt{\frac{2}{\pi n}} e^{ - \frac{m^2}{2n}}
\label{qnmcriti}
\end{eqnarray}

Let us now consider the disordered region $p_1=1-p_2<\frac{1}{2}$, where Eq \ref{laplacem}
becomes in the limit $s \to 0$
\begin{eqnarray}
\int_0^{+\infty} dn e^{-s n}  q_n^{disordered}(m)
 && 
\oppropto_{s \to 0} \frac{1}{(1-2 p_1)} 
e^{ m \left[ - \frac{s}{(1-2 p_1)} + \frac{2 p_1 (1-p_1) s^2}{(1-2 p_1)^3}\right]}
\label{qnmoffcritigene}
\end{eqnarray}
i.e. after Laplace inversion, 
one obtains that $q_n(m)$ is asymptotically a Gaussian distribution
\begin{eqnarray}
 q_n^{disordered}(m) 
\oppropto_{n \to +\infty} \frac{1}{ \sqrt{8 \pi p_1 p_2  n}} e^{ - \frac{[m-(1-2 p_1)n ]^2}{ 8 p_1 p_2 n}}
\label{qnmoffcriti}
\end{eqnarray}

The duality between $c_n^{disordered}(k)$ of Eq. \ref{solucnkfinite1d}
and $q_n^{ordered}(k)$ of Eq. \ref{solucnkfinite1dm}, between 
$c_n^{ordered}(k)$ of Eq. \ref{cnkoffcriti}
 and  $q_n^{disordered}(k)$ of Eq. \ref{qnmoffcriti},
and the identity between $c_n^{criti}(k)$ of Eq. \ref{cnkcriti}
and $q_n^{criti}(k)$ of Eq. \ref{qnmcriti}
are in agreement with the duality properties
of the one-dimensional model itself \cite{fisher}.
So even if the Boundary Strong Disorder procedure seems to break explicitly
the duality of the model by treating differently the couplings and the transverse fields,
the rules of Eqs \ref{jextnewa} and \ref{hmub} are sufficiently symmetric to
reproduce dual results as it should.

\subsection{ Dynamical exponent $z$  and Griffiths phases }

In dimension $d=1$, the dynamical exponent $z$ in the disordered phase
is known to be determined by the exact criterion \cite{exactz}
\begin{eqnarray}
1= \overline{ \left( \frac{J}{h} \right)^{\frac{1}{z}} }
\label{zexact1d}
\end{eqnarray}
For the binary distribution of Eqs \ref{binary}, \ref{slopea1} that we consider, the dynamical exponent
reads
\begin{eqnarray}
z_{exact}= \frac{ \ln \frac{J}{h_1} }{ \ln \left(\frac{1-p_1}{p_1}\right)}
\label{zexact1dres}
\end{eqnarray}
It diverges at criticality. The Griffiths phase is the region near criticality where $z>1$.
Within the exact strong disorder RG approach, the dynamical exponent appears as the coefficient
of the exponential decay of the logarithm of renormalized transverse-field $h^R$ \cite{review_strong}
\begin{eqnarray}
P(\ln h_R) \oppropto_{\ln h_R \to -\infty}  e^{ - \frac{1}{z} \vert \ln h_R \vert }
\label{probaloghR}
\end{eqnarray}

Within our boundary strong disorder renormalization, 
we have found the exponential decay of Eq. \ref{solucnkfinite1d}
for the variable $k= \frac{\ln (h^R_i/h_i)}{\ln \frac{h_1}{J}}$ (see Eq. \ref{rk}) :
this corresponds to the form of Eq. \ref{probaloghR} with the coefficient
\begin{eqnarray}
\frac{1}{z} = \frac{ \ln \left(\frac{p_1}{1-p_1}\right) }{\ln \frac{h_1}{J} }
\label{zfound}
\end{eqnarray}
that coincides with the exact value of Eq. \ref{zexact1dres}.

\subsection{ Conclusion for the one-dimensional chain $K=1$ }

These calculations for $K=1$ shows that the Boundary Strong Disorder RG procedure
is able to capture correctly for the one-dimension chain :

(i) the exact position of the critical point $p_1^{criti}=\frac{1}{2}$ (see Eq \ref{exactcriti1dbis})

(ii) the exact critical exponent $\psi=\frac{1}{2}$ governing the scaling of renormalized transverse
fields and renormalized couplings at criticality (Eqs \ref{cnkcriti} and \ref{qnmcriti})
\begin{eqnarray}
\ln h^R && \propto \ln r \propto -k  \propto - n^{\psi} \ \
 \ \ {\rm with } \ \ \psi=\frac{1}{2}
  \nonumber \\
\ln J^{ext} && \propto - m  \propto - n^{\psi}
\ \ {\rm with } \ \ \psi=\frac{1}{2}
\label{psicriti1d}
\end{eqnarray}

(iii) the exact typical correlation length exponent $\nu_{typ}=1$ governing
the divergence of the correlation length in the disordered phase
(Eq. \ref{qnmoffcriti})
\begin{eqnarray}
\overline{ \ln J^{ext}_n } && \propto - \int dm m  q_n^{disordered}(m) \propto - \frac{n}{\xi_{typ}}
 \nonumber \\
\xi_{typ} && \propto \frac{1}{1-2 p_1} \propto \left( \frac{1}{2}-p_1 \right)^{- \nu_{typ}}\
 \ \ {\rm with } \ \ \nu_{typ}=1
\label{xityp1d}
\end{eqnarray}

(iv) The dynamical exponent $z$ coincides with the exact value of the criterion of Eq. \ref{zexact1d}.

After this important check for the one-dimensional chain corresponding to $K=1$, 
we now focus on real Cayley trees having a branching ratio $K>1$.
It is convenient to consider first the simpler cases $J>h_{max}$
and $J<h_{min}$ before the more complicated case $h_{min}<J<h_{max}$, where $[h_{min},h_{max}]$
represents the support of the distribution $\pi(h_i)$ of random fields.

\section { Tree in the region $J>h_{max}$, where only bonds are decimated }

\label{sec_decibonds}

When the ferromagnetic coupling $J$ is bigger than the maximal value
$h_{max}$ of the random fields, 
it turns out that only bonds can be decimated and never sites, 
so that the RG Eq \ref{evolPK}
simplifies into
\begin{eqnarray}
P_{n+1}(J^{ext},r) && 
= \prod_{i=1}^K \int d J^{ext}_i dr_i P_{n}(J^{ext}_i,r_i) \int dh^{ini}_i \pi(h^{ini}_i )
 \delta \left(J^{ext}- \sum_{i=1}^K J^{ext}_i \right)
\delta \left (r-  \prod_{i=1}^K  \left( r_i \frac{h^{ini}_i}{J} \right)  \right)
\label{evolPKdecibondsonly}
\end{eqnarray}
i.e. all sites are included into a single quantum ferromagnetic cluster. Its coupling to the exterior reads
\begin{eqnarray}
J^{ext}_n = K^n J^{ext}_0
\label{jextdecibondsonly}
\end{eqnarray}
and its renormalized transverse-field satisfies the recurrence
\begin{eqnarray}
\ln (r_n) =\sum_{i=1}^K \left[ \ln r_{n-1}(i) + \ln \frac{h^{ini}_i}{J} \right]
\label{rdecibondsonly}
\end{eqnarray}
i.e. $ \ln (r_n)$ is the sum of 
\begin{eqnarray}
M_n \equiv K+K^2+..K^n = K \frac{K^n-1}{K-1}
\label{Mtree}
\end{eqnarray}
i.i.d variables $\left(\ln \frac{h^{ini}_i}{J}\right)$ which are all negative here :
the asymptotic distribution is thus Gaussian
\begin{eqnarray}
P_n(\ln r_n) \oppropto_{n \to +\infty} \frac{1}{\sqrt{ 2 \pi \sigma_n^2} }
 e^{- \frac{(\ln r_n- \overline{ \ln r_n})^2}{2 \sigma_n^2} } 
\label{gaussrdecibondsonly}
\end{eqnarray}
with the averaged value
\begin{eqnarray}
\overline{ \ln (r_n) }  = M_n  \overline{ \ln \frac{h^{ini}_i}{J} } \oppropto_{n \to +\infty} K^n  
 \frac{K}{K-1} \overline{ \ln \frac{h^{ini}_i}{J} }
\label{moyrdecibondsonly}
\end{eqnarray}
and with the variance
\begin{eqnarray}
\sigma_n^2 = M_n  Var \{ \ln \frac{h^{ini}_i}{J} \} \oppropto_{n \to +\infty} K^n  
 \frac{K}{K-1} Var \{ \ln \frac{h^{ini}_i}{J} \}
\label{varrdecibondsonly}
\end{eqnarray}

In conclusion for $J>h_{max}$, the tree is extremely ordered since the whole tree is a single quantum ferromagnetic
cluster, i.e. this region is very far from the transition.

\section { Tree in the region $J<h_{min}$, where only sites are decimated  }

\label{sec_decisites}

When the ferromagnetic coupling $J$ is smaller than the minimal value
$h_{min}$ of the random fields, it turns out that
 only sites are decimated and never bonds, 
so that the variable $r$ remains always $r=1$
and the RG Eq \ref{evolPK}
simplifies into
\begin{eqnarray}
P_{n}(J^{ext},r) = \delta(r-1)  P_{n}^{r=1}(J^{ext})
\label{factorr1}
\end{eqnarray}
where the distribution of the external coupling $J^{ext} $ alone evolves according to
\begin{eqnarray}
P_{n+1}^{r=1}(J^{ext})  = \prod_{i=1}^K \int d J^{ext}_i  P_{n}(J^{ext}_i) \int dh^{ini}_i \pi(h^{ini}_i )
\delta \left(J^{ext}- \sum_{i=1}^K  J^{ext}_i \frac{J}{h^{ini}_i } \right)
\label{evolPKDP}
\end{eqnarray}

\subsection{ Quantitative mapping onto a Directed Polymer on the Cayley tree}

The evolution of Eq. \ref{evolPKDP}  
coincides with the evolution with the length $L$
of the partition function $Z^{DP}_{L}(\beta)$ 
of a Directed Polymer on the Cayley tree at 'inverse-temperature' $\beta=1$
\begin{eqnarray}
  J_L^{ext}=Z_L^{DP}(\beta=1)
\label{jlzdp}
\end{eqnarray}
where the Directed Polymer model
 \begin{eqnarray}
Z^{DP}_{L}(\beta) = \displaystyle \sum_{RW} 
\exp \left( - \beta \displaystyle \sum_{1 \leq n \leq L} 
\epsilon (n, i(n))  \right) 
\label{directed}
  \end{eqnarray}
contains effective sites random energies 
given by
\begin{eqnarray}
\epsilon(n, i(n) ) = \ln h^{ini}_n(i) - \ln J
\label{epssite}
\end{eqnarray}

\subsection{ Reminder on the Directed Polymer on the Cayley tree}

The Derrida-Spohn solution for the Directed Polymer on the Cayley tree \cite{Der_Spo}
yields that the extensive term of the partition function of Eq. \ref{directed}
is given by
 \begin{eqnarray}
\ln Z^{DP}_{L}(\beta)  && \oppropto_{L \to +\infty}  - \beta f(\beta) L +...
\ \ {\rm if } \ \  \beta<\beta_c \nonumber \\
\ln Z^{DP}_{L}(\beta)  && \oppropto_{L \to +\infty}  - \beta f(\beta_c) L +...
\ \ {\rm if } \ \  \beta> \beta_c
\label{directedextensif}
  \end{eqnarray}
in terms of the following function defined on $\beta \in ]0,+\infty[$
 \begin{eqnarray}
 f(\beta)  \equiv - 
\frac{1}{\beta} \ln \left[ K J^{\beta} \int dh h^{-\beta} P(h) \right] 
\label{fbeta}
  \end{eqnarray}
and where the location $\beta_c$  of the freezing critical point is given by the condition 
 \begin{eqnarray}
 \partial_{\beta} f(\beta) \vert_{\beta=\beta_c} =0 
\label{fprimebeta}
  \end{eqnarray}

In the limit $\beta \to 0$, the normalization $\int dh P(h)=1$
leads to the expansion
 \begin{eqnarray}
 f(\beta)  \oppropto_{\beta \to 0}  - \ln J - \frac{\ln K }{\beta} \to - \infty
\label{fbetazero}
  \end{eqnarray}

In the limit  $\beta \to +\infty$, we have to distinguish two cases 

(i) if the distribution $P(h)$ is continuous, one may perform a saddle-point calculation at $h_{min}$
with the following change of variables $h=h_{min}(1+x/\beta)$ to obtain
 \begin{eqnarray}
\int_{h_{min}}^{..} dh h^{-\beta} P(h) 
\simeq \int_{0}^{+\infty} h_{min} \frac{dx}{\beta} h_{min}^{-\beta}
e^{- x } P(h_{min}) \simeq \frac{h_{min}^{1-\beta}  P(h_{min}) }{\beta}
\label{colhmin}
  \end{eqnarray}
that leads to the following expansion
 \begin{eqnarray}
 f(\beta)  && \oppropto_{\beta \to +\infty}  - \ln J - \frac{\ln K }{\beta} 
- \frac{1}{\beta} \ln \left[ \frac{h_{min}^{1-\beta}  P(h_{min}) }{\beta}  \right] 
=
 - \ln J - \frac{\ln K }{\beta} 
- \frac{1}{\beta} \left[ (1-\beta) \ln h_{min} + \ln   P(h_{min}) -\ln \beta \right]
 \nonumber \\
&& \oppropto_{\beta \to +\infty} - \ln J + \ln h_{min} + \frac{\ln \beta}{\beta} - \frac{\ln K +\ln h_{min} + \ln   P(h_{min}) }{\beta} 
\label{fbetainfinitydv}
  \end{eqnarray}
The positivity of the leading correction
 $\frac{\ln \beta}{\beta} >0$ shows that the positive limit
 \begin{eqnarray}
 f(\beta=+\infty) = - \ln J + \ln h_{min} >0
\label{fbetainfinityvalue}
  \end{eqnarray}
is reached from above.
Between these two limits, the maximum reached at $\beta_c$ is thus positive
 \begin{eqnarray}
 f(\beta_c) >   f(\beta=+\infty) >0
\label{fbetac}
  \end{eqnarray}

(ii) if the distribution $P(h)$ is the binary distribution of Eq. \ref{binary}
 \begin{eqnarray}
\int dh h^{-\beta} P(h) = p_1 h_1^{-\beta} \left( 1 + \frac{ p_2 h_1^{\beta} h_2^{-\beta} }{ p_1  } \right)
\label{colhminbinary}
  \end{eqnarray}
then the asymptotic behavior reads
 \begin{eqnarray}
 f(\beta)  && \oppropto_{\beta \to +\infty} \ln h_1  - \ln J - \frac{\ln (K p_1)}{\beta} 
 - \frac{1}{\beta}  \frac{ p_2 h_1^{\beta} h_2^{-\beta} }{ p_1  }  
\label{fbetainfinitybinarydv}
  \end{eqnarray}
So the positive limit 
 \begin{eqnarray}
 f(\beta=+\infty) = - \ln J + \ln h_1 >0
\label{fbetainfinitybinaryvalue}
  \end{eqnarray}
will be reached from above if $K p_1 <1 $ and from below if $K p_1 >1 $.
For the case $K p_1 <1 $, one has thus a finite location $\beta_c$ satisfying Eq \ref{fbetac}.
For the case $K p_1 >1 $, there is no freezing, i.e. in the limit of zero temperature $\beta \to +\infty$, the DP is able to find a path containing only $h_1$ (since $p_1$ is above the percolation transition $p_1>p_1^{perco}=1/K$).

\subsection{ Application to the variable $J^{ext}$ of the quantum model }

The solution of Eq. \ref{directedextensif} can be directly translated
for the external coupling $ J_L^{ext} $ via the mapping of Eq. \ref{jlzdp}
 \begin{eqnarray}
\ln J_L^{ext} =  \ln Z^{DP}_{L}(\beta=1)  && \oppropto_{L \to +\infty}  -  f(1) L +...
\ \ {\rm if } \ \  1<\beta_c \nonumber \\
\ln J_L^{ext} = \ln Z^{DP}_{L}(\beta=1)  && \oppropto_{L \to +\infty}  -  f(\beta_c) L +...
\ \ {\rm if } \ \  1> \beta_c
\label{directedextensifq}
  \end{eqnarray}

For the case $1> \beta_c$, the inequality $f(\beta_c) >0$ shows that
$\ln J_L^{ext} $ decays with $L$ so that the quantum model can then only be
in its disordered phase.

For the case $1< \beta_c$, we see that the quantum model can be either disordered if $ f(1)>0$,
 ordered if $f(1)<0$ and at criticality if $f(1)=0$.
This condition for criticality reads using Eq \ref{fbeta}
 \begin{eqnarray}
1 = K J_c \int dh h^{-1} P(h)  
\label{jcquantdeloc}
  \end{eqnarray}
The condition $1< \beta_c$ to be in the delocalized phase of the Directed Polymer
is equivalent to the condition 
 \begin{eqnarray}
0<f'(\beta=1) =    \ln \left[ K  \int dh h^{-1} P(h) \right] 
+ \frac{  \int dh h^{-1} ( \ln h) P(h) }{\int dh h^{-1} P(h)}
\label{conditionfprimebeta}
  \end{eqnarray}
Using the transition condition $0= f(\beta=1)  = -   \ln \left[ K J_c \int dh h^{-1} P(h) \right]  $,
we may rewrite the condition as
 \begin{eqnarray}
0< f'(\beta=1)  =    - \ln J_c
+ \frac{  \int dh h^{-1} ( \ln h) P(h) }{\int dh h^{-1} P(h)} = 
\frac{  \int dh h^{-1} ( \ln h- \ln J_c) P(h) }{\int dh h^{-1} P(h)}
\label{fprimebeta1bis}
  \end{eqnarray}
which is always satisfied if $J_c<h_{min}$, which is necessary 
to have the quantitative mapping onto the DP (see Eq. \ref{evolPKDP}).

\subsection{ Conclusion  }

The conclusion of this analysis of the region $0<J<h_{min}$,
where there exists a quantitative mapping onto a Directed Polymer on the Cayley tree
is the following :

(i) either the transition of the quantum model takes place in this region $0<J_c \leq h_{min}$.
Then this can only correspond to the delocalized phase of the Directed Polymer,
where the external coupling has for asymptotic behavior around $J_c$
 \begin{eqnarray}
J_L^{ext} = Z^{DP}_{L}(\beta=1)  \propto e^{  -  f(1) L } 
= \left[K J \int dh h^{-1} P(h) \right]^L = \left[\frac{J}{J_c} \right]^L
\label{jldpdeloc}
  \end{eqnarray}
In particular, in the disordered phase $J<J_c$, the typical external coupling decays
exponentially
 \begin{eqnarray}
\ln J_L^{ext} \propto - \frac{L}{\xi_{typ}} 
\label{jldpdelocdes}
  \end{eqnarray}
where the correlation length $\xi_{typ}$ diverges as
 \begin{eqnarray}
\xi_{typ} = \frac{1}{ \ln \left(\frac{J_c}{J} \right) } \propto (J_c-J)^{-\nu_{typ}} \ \ 
{\rm with } \ \ \nu_{typ}=1
\label{jldpdelocxityp}
  \end{eqnarray}
Note that this scenario where the ordered phase appeared at $J_c<h_{min}$ where
one decimates only sites and no bonds, i.e. where no clusters are formed
is completely different from what happens in $d=1$ (see section \ref{sec_1d}),
and is due to the branching with ratio $K>1$ of the tree.

(ii) or the quantum model remains disordered in the whole region $0<J \leq h_{min}$,
i.e. $J_L^{ext} $ decays exponentially in $L$ even for $J=h_{min}$.
Then the transition will take place in the region $h_{min}<J<h_{max}$ where 
one has to decimate both sites and links.
This is the case studied in  section \ref{sec_deciboth}.

\subsection{ Special case of the binary distribution  }

For the binary distribution of Eq. \ref{binary} 
discussed around Eq. \ref{fbetainfinitybinaryvalue},
various cases can appear :

(a) For the case $ p_1 > \frac{1}{K} $, there is no freezing at finite $\beta_c$ 
for the corresponding Directed Polymer model, so that the behavior of the external coupling 
is given by Eq. \ref{jldpdeloc}
in the whole region $0<J \leq h_{min}=h_1$ 
 \begin{eqnarray}
J_L^{ext} && = Z^{DP}_{L}(\beta=1)  \propto e^{  -  f(1) L } = \left[\frac{J}{J_c} \right]^L 
  \end{eqnarray}
where
 \begin{eqnarray}
J_c && = \frac{1}{K  \int dh h^{-1} P(h) } =  \frac{1}{K  \left[ \frac{p_1}{h_1}+ \frac{p_2}{h_2} \right] } \leq  \frac{ h_1}{ (K p_1) } < h_1
\label{jcbinary}
  \end{eqnarray}
i.e. the quantum transition always takes place at $J_c<h_{min}=h_1$, where there exists the quantitative mapping
onto a Directed Polymer model in its delocalized phase.

(b) For the case $p_1 < \frac{1}{K} $, there exists a freezing transition for the associated Directed Polymer model at some finite $\beta_c$, and one has to discuss whether $\beta_c>1$ or $\beta_c<1$.

According to the discussion of section \ref{sec_pure}, a pure system
 with the uniform random field $h_0$ taking the value $h_{max}$
would have a pure transition at $J_c^{pure}(h_{max})=\frac{h_{max} }{K} $,
so one expects a lower critical coupling for the disordered
 case when some random field are smaller 
$J_c< J_c^{pure}(h_{max})=\frac{h_{max} }{K}$.
So if one chooses $\frac{h_{max} }{K} < h_{min}$, one is sure to have $J<h_{min}$, i.e. the transition
will takes place
 in the region where only sites are decimated (section \ref{sec_decisites}) 
and where there exists the quantitative mapping onto the Directed Polymer described above. 

On the other hand, if one chooses a binary distribution $(h_1<h_2)$ with $p_1=1-p_2 \to 0$,
then the transition will be near $J_c \simeq J_c^{pure}(h_2)=\frac{h_2 }{K} $,
so if one chooses $h_1<\frac{h_2 }{K}$, the transition should take place in the region $h_1<J<h_2$
where one needs to take into account the formation of quantum ferromagnetic clusters,
that we discuss in the following section.

\section { Tree in the region $h_{min}<J<h_{max}$ where both sites and links are decimated  }

\label{sec_deciboth}

In this section, we consider the remaining case  $h_{min}<J<h_{max}$ where both sites and links are decimated.

\subsection { Evolution of the distribution of the variable $r$ }

Let us first focus on the distribution of the renormalized transverse fields alone,
i.e. on the distribution $P_n(r)$
\begin{eqnarray}
P_n(r) \equiv \int dJ^{ext} P_n(J^{ext},r)
\label{ralonebis}
\end{eqnarray}
that evolves according to the recurrence (after integrating Eq \ref{evolPK}  over $J^{ext}$)
\begin{eqnarray}
P_{n+1}(r) && = \prod_{i=1}^K \int dr_i P_{n}(r_i) 
\int  d r^{new}_i \int dh^{ini}_i \pi(h^{ini}_i )
\left[ \theta( h^{ini}_i r_i \geq J) 
\delta(r^{new}_i -1  ) 
+ \theta( h^{ini}_i r_i <J) \delta(r^{new}_i - r_i \frac{h^{ini}_i}{J}  )  \right]
\nonumber \\
&& \delta \left (r-  \prod_{i=1}^K  r^{new}_i  \right) 
\label{evolPKr}
\end{eqnarray}
with the initial condition
\begin{eqnarray}
P_{n=0}(r)  = \delta(r-1)
\label{initialPKr}
\end{eqnarray}

\subsection { Special case of the binary distribution }

To simplify the technical details, we now focus again on the binary distribution of Eq. \ref{binary}
and we fix $h_1<J=\sqrt{h_1 h_2}<h_2$ so that Eq \ref{slopea1} holds.
The control parameter of the transition is the fraction $p_1=1-p_2$ of Eq. \ref{binary}.

Starting from the initial condition of Eq \ref{initialPKr},
the only values that can appear are of the form of Eq. \ref{rk}
i.e. the probability distribution $P_n(r)$ takes the form of a sum of delta functions
\begin{eqnarray}
P_n(r) = \sum_{k=0}^{+\infty} c_n(k)  \delta \left[ r- \left( \frac{h_1}{J}  \right)^{k}  \right]
\label{ansatzk}
\end{eqnarray}
with the normalization
\begin{eqnarray}
1= \int_0^1 dr P_n(r) =  \sum_{k=0} c_n(k) 
\label{norman}
\end{eqnarray}
and the initial condition (Eq \ref{initialPKr})
\begin{eqnarray}
 c_{n=0}(k) = \delta_{k,0}
\label{initialck}
\end{eqnarray}

Plugging the form of Eq. \ref{ansatzk} into the RG Equation of Eq. \ref{evolPKr}
yields that the generation function
\begin{eqnarray}
{\hat c}_n(z) \equiv  \sum_{k=0}^{+\infty}  z^k c_n(k) 
\label{defcz}
\end{eqnarray}
satisfies the recurrence
\begin{eqnarray}
 {\hat c}_{n+1} (z)  && = \left( {\hat c}_{n}(z) \left[  p_1  z +  \frac{p_2}{z}  \right]
+ p_2 {\hat c}_{n} (0) \left[ 1  -  \frac{1}{z}   \right] \right)^K
\label{recgenecK}
\end{eqnarray}
with the initial condition  (eq \ref{initialcmk})
\begin{eqnarray}
  {\hat c}_{n=0} (z) = 1
\label{hatCini}
\end{eqnarray}

In one dimension $K=1$, where the corresponding equation was linear
and could be solved for any $n$ via the introduction of the time generation function
 (Eq \ref{recgenec1dC}).
Here for $K>1$, the equation is not linear anymore, and we do not know
how to solve exactly Eq. \ref{recgenecK} for any finite size $n$.
So the goal will be here to determine the possible
asymptotic behaviors of the distribution $c_n(k)$ as $n \to +\infty$,
 as a function of the parameter $p_1=1-p_2 \in ]0,1[$.

\subsection { First discussion on the possible behaviors as $n \to +\infty$  }

It is clear that the asymptotic behaviors as $n \to +\infty$
will depend on the behavior of $c_n(k)$ near the origin $k=0,1,2,..$
and on the behaviors of the first moments.
It is thus convenient to write the corresponding recurrences.

\subsubsection { Recurrences for $c_n(k)$ near the origin $k=0,1,2$   }

From Eq. \ref{recgenecK}, one obtains that the recurrences for $c_n(k)$
for the first values of $k=0,1,2$ reads
\begin{eqnarray}
  c_{n+1} (0) && = \left[ p_2 (c_n(0)+c_n(1)) \right]^K  \label{k012}
 \\   
 c_{n+1} (1) && = K \left[ p_2 (c_n(0)+c_n(1)) \right]^{K-1} \left[p_1 c_n(0)+p_2 c_n(2) \right]
\nonumber \\   \nonumber
 c_{n+1} (2) && = K \left[ p_2 (c_n(0)+c_n(1)) \right]^{K-1} \left[ p_1 c_n(1)+p_2 c_n(3) \right]
+ \frac{K (K-1)}{2}\left[ p_2 (c_n(0)+c_n(1)) \right]^{K-2}\left[p_1 c_n(0)+p_2 c_n(2) \right]^2
\end{eqnarray}

In the limit $n \to +\infty$, there are only two possibilities :

(i) either $c_n(k)$ converge towards finite values $c_{\infty}(k) $
\begin{eqnarray}
  c_{n} (k=0,1,2,..) && \oppropto_{n \to +\infty} c_{\infty} (k=0,1,2,..) >0
\label{cnasymp}
\end{eqnarray}
that satisfy Eqs \ref{k012} as stationary equations.

(ii) or $c_n(k)$ for finite values $k=0,1,2,..$ converge towards zero as $n \to +\infty$ :
then from Eq. \ref{k012}, the only way they can decay to zero is exponentially with respect to $K^n$
\begin{eqnarray}
  c_{n} (k=0,1,2,..) && \oppropto_{n \to +\infty}  e^{- (cst) K^n}
\label{cnasympexp}
\end{eqnarray}

\subsubsection { Recurrences for the first moment $<k>_n$   }

Using the series expansion in the moments
\begin{eqnarray}
{\hat c}_n(z=e^{-q}) && = \sum_{k=0}^{+\infty} e^{-q k} c_k 
= \sum_{k=0}^{+\infty} \left(1-q k + \frac{q^2 k^2}{2}
+O(q^3) \right)  c_k
=  1 - q <k>_n + \frac{q^2}{2} <k^2>_n +O(q^3)
\label{dvq}
\end{eqnarray}
one may derive from Eq. \ref{recgenecK}
 the recurrences for the first moment
\begin{eqnarray}
 <k>_{n+1} =  K \left[ <k>_n +(2 p_1- 1) + p_2 c_{n} (k=0) \right]
\label{rec1ermoment}
\end{eqnarray}

In the limit $n \to +\infty$, there are only two possibilities :

(a) either $<k>_n$ converge towards a finite value $<k>_{\infty} $
that satisfies Eq \ref{rec1ermoment} as stationary equation,
where $c_{\infty} (k=0) $ has to remain finite
\begin{eqnarray}
 <k>_{\infty} =  K \left[ <k>_{\infty} +(2 p_1- 1) + p_2 c_{\infty} (k=0) \right]
\label{rec1ermomentinfty}
\end{eqnarray}
For $K=1$, this fixes the value $c_{\infty} (k=0) = (1-2 p_1)/p_2 $ found previously 
(Eq \ref{solucnkfinite1d})
whereas for $K>1$, it yields 
\begin{eqnarray}
 <k>_{\infty} =  \frac{K }{K-1} \left[ (1-2 p_1) - p_2 c_{\infty} (k=0) \right] 
\label{rec1ermomentinftyres}
\end{eqnarray}
which should be positive, since $k$ is a positive random variable,
so we have the bound
\begin{eqnarray}
0 <  c_{\infty} (k=0) \leq \frac{ (1-2 p_1)}{p_2}   \ \ {\rm implying } \ \ p_1 < \frac{1}{2}
\label{bound1}
\end{eqnarray}
In fact, since $k$ is an integer random variable, we have in fact the stronger constraint
\begin{eqnarray}
 <k>_{\infty} = \sum_{k=1}^{+\infty} k  c_{\infty} (k) \geq \sum_{k=1}^{+\infty}   c_{\infty} (k) =1-c_{\infty} (k=0)
\label{moykinfty}
\end{eqnarray}
which yields with Eq. \ref{rec1ermomentinftyres}
\begin{eqnarray}
(1- 2 K p_1) - (1-K p_1) c_{\infty} (k=0)\geq 0
\label{bound2}
\end{eqnarray}
For instance for $K=2$, this yields
\begin{eqnarray}
 c_{\infty}^{(K=2)} (k=0)\leq \frac{1-4 p_1}{1-2 p_1}  \ \ {\rm implying } \ \ p_1 < \frac{1}{4}
\label{bound2k2}
\end{eqnarray}

(b) or $<k>_n$ diverges as $n \to +\infty$ :
then Eq. \ref{rec1ermoment} can be approximated by $ <k>_{n+1} \simeq  K  <k>_n $ 
so the only way it can diverge is like
\begin{eqnarray}
 <k>_n \oppropto_{n \to +\infty } K^n
\label{knavasymp}
\end{eqnarray}

\subsubsection { Recurrences for the variance $\sigma_n^2 \equiv <k^2>_n-<k>_n$  }

From the recurrence concerning the second moment
\begin{eqnarray}
 <k^2>_{n+1}
=  && K [  <k^2>_n +(K-1) <k>_n^2 + 2 \left[ K(2p_1-1)+(K-1)p_2 c_n(k=0) \right] <k>_n 
\nonumber \\   &&
+  K(2p_1-1)^2 + 4  p_1 p_2 + p_2 \left(2K(2p_1-1) +1-4 p_1 \right)c_n(k=0) + (K-1) p_2^2c_n^2(k=0) ]
\label{rec2dmoment}
\end{eqnarray}
one obtains that the variance
\begin{eqnarray}
\sigma_n^2 \equiv  <k^2>_n-   <k>_{n}^2 
\label{defsigma}
\end{eqnarray}
satisfies the recurrence 
\begin{eqnarray}
\sigma_{n+1}^2  
&& =  K \left[ \sigma_n^2 - 2 p_2 c_n(k=0)  <k>_{n}
+ 4 p_1p_2 + (1 - 4 p_1)p_2 c_n(k=0) -p_2^2 c_n^2(k=0)
\right]
\label{recsigma}
\end{eqnarray}

In the limit $n \to +\infty$, there are only two possibilities :

(a) either $\sigma_n^2$ converge towards a finite value $\sigma_{\infty}^2 $
should be stable by the recurrence of Eq. \ref{recsigma}
\begin{eqnarray}
\sigma_{\infty}^2  
&& =  K \left[ \sigma_{\infty}^2 - 2 p_2 c_{\infty}(k=0)  <k>_{\infty}
+ 4 p_1p_2 + (1 - 4 p_1)p_2 c_{\infty}(k=0) -p_2^2 c_{\infty}^2(k=0)
\right]
\label{recsigmainfty}
\end{eqnarray}
i.e. using Eq. \ref{rec1ermomentinftyres}, this yields
\begin{eqnarray}
\sigma_{\infty}^2  
&& =  \frac{K}{(K-1)^2} p_2
 \left[ - p_2 (K+1) c_{\infty}^2(k=0)
+ (K+1-4 p_1) c_{\infty}(k=0)
- 4 p_1 (K-1)
\right]
\label{recsigmainftysol}
\end{eqnarray}
The positivity of the variance implies that the discriminant of the right-handside second order polynomial should be positive, and that $c_{\infty}(k=0) \in ]c_- , c_+ [$ 
where $c_{\pm}$ are the two real roots.

For instance for $K=2$, Eq. \ref{recsigmainftysol} becomes
\begin{eqnarray}
\sigma_{\infty}^2  
&& =  2 p_2
 \left[ - 3 p_2  c_{\infty}^2(k=0) + (3-4 p_1) c_{\infty}(k=0) - 4 p_1 
\right]
\label{recsigmainftysolk2}
\end{eqnarray}
with the discriminant 
\begin{eqnarray}
 \delta=9  -72 p_1+64 p_1^2 
\label{recsigmainftysolk2delta}
\end{eqnarray}
Since $0 \leq p_1 < 1/2$, this discriminant is positive only if
\begin{eqnarray}
 p_1 < p_1^{max} = \frac{ 3 (3-\sqrt{5})}{16} = 0.143237
\label{recsigmainftysolk2deltabound}
\end{eqnarray}

(b) or $\sigma_n^2$ diverges as $n \to +\infty$ :
then from Eq. \ref{recsigma}, the only way it can diverge is like
\begin{eqnarray}
 \sigma_n^2 \oppropto_{n \to +\infty } K^n
\label{sigma2asymp}
\end{eqnarray}

\subsubsection { Summary  }

From the possible behaviors of $c_n(k=0)$ and of the two first moments as $n \to \infty$,
we have seen that either a finite $c_n(k)$ exists with no rescaling in $k$,
but this can only be the case for sufficiently small $p_1$, or the variable $k$ flows towards infinity
with the first moment and the variance growing as $K^n$, as we already found for $J>h_{max}$
in section \ref{sec_decibonds}. Let us first discuss the case $p_1>\frac{1}{2}$
where the solution found in  section \ref{sec_decibonds} can be simply extended.

 \subsection{ Asymptotic rescaled distribution $c_{\infty}(k)$  for $p_1>\frac{1}{2}$  }

For $p_1=1$, one has $J>h_{max}=h_1$ and one should recover the solution of section \ref{sec_decibonds}.
It is clear that this type of solution can survive in the whole region $p_1>\frac{1}{2}$
for the following reasons.
Let us assume that ${\hat c}_{n} (0) $ is negligible in Eq. \ref{recgenecK} :
then, taking into account the initial condition of Eq. \ref{hatCini},
 one has the simple solution in terms of $M_n$ of Eq. \ref{Mtree}
\begin{eqnarray}
 {\hat c}_{n} (z)  && = \left( {\hat c}_{n-1}(z) \left[  p_1  z +  \frac{p_2}{z}  \right]
    \right)^K =  \left( {\hat c}_{0}(z) \right)^{K^n} \left[  p_1  z +  \frac{p_2}{z}  \right]^{M_n}
 =  \left(   p_1  z +  \frac{p_2}{z}   \right)^{M_n}
= \sum_{j=0}^{M_n} C_{M_n}^j \left[ p_1  z \right]^j \left[    \frac{p_2}{z}  \right]^{M_n-j} 
\label{cfree}
\end{eqnarray}
i.e. the coefficient of order $k$ simply reads (Eq \ref{defcz})
\begin{eqnarray}
c_n(k) = C_{M_n}^{\frac{M_n+k}{2}}  p_1^{\frac{M_n+k}{2}}
  p_2^{\frac{M_n-k}{2}}
\label{cnkfree}
\end{eqnarray}
For $n \to +\infty$, we may use the Stirling formula to obtain the asymptotic Gaussian formula
\begin{eqnarray}
c_n(k) \oppropto_{n \to +\infty} \frac{1}{ \sqrt{8 \pi p_1 p_2  M_n}} e^{ - \frac{[k-(2 p_1-1)M_n ]^2}{ 8 p_1 p_2 M_n}}
\label{cnkfreegauss}
\end{eqnarray}
that coincides with Eq. \ref{cnkoffcriti} for the one dimensional chain $K=1$ where $M_n=n$.
This solution is consistent as long as 
\begin{eqnarray}
c_n(k=0) \oppropto_{n \to +\infty} \frac{1}{ \sqrt{8 \pi p_1 p_2  M_n}} e^{ - M_n \frac{(2 p_1-1)^2}{ 8 p_1 p_2 }}
\label{cnkfreegaussk0}
\end{eqnarray}
remains exponentially negligible in $M_n$ with respect to the center of the distribution localed around
$ k \sim (2 p_1-1)M_n $ i.e. for $p_1>\frac{1}{2}$.

 \subsection{ Existence of a finite asymptotic distribution $c_{\infty}(k)$  as $n \to \infty$  for small enough $p_1$ }

If a finite asymptotic distribution $c_{\infty}(k)$ exists as $n \to +\infty$ without any rescaling in $k$,
it should satisfy the stationary version of the recurrence of Eq. \ref{recgenecK}
\begin{eqnarray}
z^K {\hat c}_{\infty} (z)  && = \left( {\hat c}_{\infty}(z) \left[  p_1  z^2 +  p_2  \right]
+ p_2 {\hat c}_{\infty} (0) \left[ z  -  1   \right] \right)^K
\label{recgenecKstatio}
\end{eqnarray}

Note that for $p_1=1-p_2=0$, we know that we have the trivial solution 
\begin{eqnarray}
 {\hat c}_{\infty}^{(p_1=0)}(z) =1 \ \ \ \ \ \ { \rm i.e.  } \ \ \ \ \ c_{\infty}^{(p_1=0)}(k) = \delta_{k,0}
\label{solup2eq1}
\end{eqnarray}
A stationary solution $c_{\infty}(k) $ is thus likely to exist for small enough $p_1=1-p_2$.

Let us now focus on the case of branching ratio $K=2$, where this equation is simply 
quadratic
\begin{eqnarray}
\left[  p_1 z^2 + p_2 \right]^2 {\hat c}_{\infty}^2( z) 
+   \left( 2  p_2 {\hat c}_{\infty} (0) \left[  p_1 z^2 + p_2 \right]
 \left[ z -1  \right]- z^2 \right){\hat c}_{\infty}( z)
+ p_2^2 {\hat c}_{\infty}^2 (0) \left[ z -1  \right]^2  =0
\label{recgenec1dCK2star}
\end{eqnarray}
and involves the discriminant
\begin{eqnarray}
\Delta(z) && = z^2   4  p_2^2 {\hat c}_{\infty} (0) R(z)
\nonumber \\
R(z) && =    \left[ 1+\frac{ p_1}{p_2} z^2 \right]
 \left[ 1-z  \right] + \frac{z^2}{4  p_2^2 {\hat c}_{\infty} (0)} 
\label{discriK2}
\end{eqnarray}
The two possibles solutions read
\begin{eqnarray}
{\hat c}_{\infty}^{\pm}( z) = \frac{  \left( 2  p_2 {\hat c}_{\infty} (0) \left[  p_1 z^2 + p_2 \right]
 \left[ 1 -z \right] + z^2 \right) \pm \sqrt{\Delta(z) } }{ 2 \left[  p_1 z^2 + p_2 \right]^2}
\label{soluCKstar}
\end{eqnarray}
At $z=0$, both solutions coincide with ${\hat c}_{\infty} (0) $.
At $z=1$, the value $\Delta(z=1)=1$ yields
\begin{eqnarray}
{\hat c}_{\infty}^{\pm}( z=1) = \frac{ 1 \pm 1  }{ 2 }
\label{soluCKstar1}
\end{eqnarray}
So the solution with the proper normalization $1= {\hat c}_{\infty}( z=1)$
is the solution with the $(+)$ sign
\begin{eqnarray}
{\hat c}_{\infty}( z) ={\hat c}_{\infty}^{+}( z)
 && = {\hat c}_{\infty}( 0)
\frac{ R(z) + \frac{z^2}{4  p_2^2 {\hat c}_{\infty} (0)} 
+ \sqrt{ \frac{z^2}{p_2^2 {\hat c}_{\infty} (0) } R(z) } }{\left[ 1+ \frac{p_1}{p_2} z^2  \right]^2}
 = {\hat c}_{\infty}( 0)
\frac{\left[ 1 -z  \right]^2}{ R(z) + \frac{z^2}{4  p_2^2 {\hat c}_{\infty} (0)} 
- \sqrt{ \frac{z^2}{p_2^2 {\hat c}_{\infty} (0) } R(z) } }
\label{soluCKstarplus}
\end{eqnarray}
Since $R(z)$ of Eq. \ref{discriK2} is polynomial of third degree having $R(z=0)=1$,
these expressions shows that Eq \ref{soluCKstarplus} can be expanded in power-series of $z$,
and the coefficients $c_{\infty}(k)$ can be computed via the following contour integral in the complex plane
\begin{eqnarray}
c_{\infty}(k) && = \oint_{C_0} \frac{d z }{2i \pi z^{k+1} } {\hat c}_{\infty}( z)
\label{invcsatrK}
\end{eqnarray}
Now we have to impose that these coefficients $c_n(k)$ that represent probabilities are all
real and positive.
So we have to discuss the cuts that appear in $\sqrt{R(z)}$
i.e. the three roots $z_i$ of the cubic polynomial
 $R(z)$ of Eq. \ref{discriK2} that can be written as 
\begin{eqnarray}
R(z) = a z^3+b z^2+c z +d = \frac{p_1}{p_2} \prod_{i=1}^3 (z_i-z)
\label{Rz}
\end{eqnarray}
with parameters
\begin{eqnarray}
d && =1
\nonumber \\
c && =-1
\nonumber \\
b && = \left[\frac{p_1}{p_2} + \frac{1}{4  p_2^2 {\hat c}_{\infty} (0)}\right] 
\nonumber \\
a && = - \frac{p_1}{p_2} 
\label{polynomeR}
\end{eqnarray}
To determine the three roots $z_i$ of Eq. \ref{Rz}, it is convenient to perform the following change of variables
\begin{eqnarray}
z=x-\frac{b}{3a} = x +
 \frac{\left[1 + \frac{1}{4 p_1 p_2 {\hat c}_{\infty} (0)}\right] }{3  }
\label{zx}
\end{eqnarray}
so that the new variable $x$ satisfies the standard form of cubic equation
\begin{eqnarray}
x^3+Px+Q =0
\label{eqx}
\end{eqnarray}
of parameters
\begin{eqnarray}
P && = \frac{c}{a}- \frac{b^2}{3a^2}
=- \frac{\left[ 1+8 p_1 p_2{\hat c}_{\infty} (0) + 16 p_1 p_2^2 (4p_1-3){\hat c}_{\infty}^2 (0) \right]  }
{48 p_1^2 p_2^2 {\hat c}_{\infty}^2 (0)}
\nonumber \\
Q && = \frac{d}{a} + \frac{b}{ 27 a} \left[\frac{2 b^2}{ a^2}- \frac{9c}{  a}\right]
= -  \frac{\left[1 +12 p_1 p_2 {\hat c}_{\infty} (0)+24 p_1 p_2^2 (5 p_1-3) {\hat c}_{\infty}^2 (0)
+64 p_1^2 p_2^3 (9-8 p_1) {\hat c}_{\infty}^3 (0)
\right]}{864 p_1^3 p_2^3 {\hat c}_{\infty}^3 (0)} 
\label{eqxPQ}
\end{eqnarray}
The discriminant of this cubic equation reads
\begin{eqnarray}
\Delta_3 && \equiv Q^2+ \frac{4}{27}P^3 = \frac{N_3}{432 p_1^4 p_2^2 {\hat c}_{\infty}^3 (0)} \nonumber \\
N_3 && = 1 + p_2( 13 p_1-1){\hat c}_{\infty} (0) +
 16 p_2^2 p_1 (8p_1-5) {\hat c}_{\infty}^2 (0)+ 64p_1 p_2^3{\hat c}_{\infty}^3 (0)
\label{Delta3}
\end{eqnarray}
The numerator $N_3$ is itself a third-degree polynomial in the variable ${\hat c}_{\infty}^3 (0)$
: again it is convenient to perform the change of variables
\begin{eqnarray}
{\hat c}_{\infty}(0)=t + \frac{2}{3} - \frac{1}{4 p_2}
\label{ct}
\end{eqnarray}
to rewrite the numerator in the reduced form
\begin{eqnarray}
N_3=64 p_1 p_2^3  (t^3+pt+q )
\label{n3t}
\end{eqnarray}
of parameters
\begin{eqnarray}
p && = - \frac{ 3 + 64 p_1 (1-4 p_1)}{192 p_1 p_2}
\nonumber \\
q && = \frac{63-280 p_1+3584p_1^2-4096 p_1^3 }{6912 p_1 p_2^2 }
\label{pqDelta3n3}
\end{eqnarray}
The corresponding discriminant reads
\begin{eqnarray}
\delta_3 && =  q^2+ \frac{4}{27}p^3 = -  \frac{(1 - 28 p_1)^3}{1769472 p_1^3  p_2^4 } 
\label{Delta3n3}
\end{eqnarray}

The possible solutions $z_i$ of the cubic polynomial $R(z)$ of Eq. \ref{Rz}
depend on the sign of the polynomial $\Delta_3$ of Eq. \ref{Delta3}

(i) If $\Delta_3>0$, there exists a single real root, and two complex roots that are complex conjugates.

(ii) If $\Delta_3<0$, the three roots $z_i$ are real.

(iii) If $\Delta_3=0$, there exists a simple real solution $z_s$ and a double real solution $z_d $

Moreover, since
the generating function $c(z)$ contains only non-negative coefficients $c(k) \geq 0$,
 we known from Pringsheim theorem that its closest singularity to the origin
has to be on the real axis.

In the limit $p_1 \to 0$ where we know the solution (Eq. \ref{solup2eq1}),
 $R(z)$ reduces to the second order polynomial 
\begin{eqnarray}
R^{(p_1=0)}(z)=\frac{(2-z)^2}{4}
\label{rzp1zero}
\end{eqnarray}
corresponding to the limiting case where the double real solution is finite,
and the simple real root diverges
\begin{eqnarray}
z_d^{(p_1=0)} && =2  \nonumber \\
z_s^{(p_1=0)} && \to +\infty
\label{zdzsp1zero}
\end{eqnarray}

By continuity in $p_1$, we expect that $z_s$ will become finite but large.
The double real solution could in principle separate,
but they cannot become complex (case  $\Delta_3>0$) as a consequence of Pringsheim theorem,
since they are closest to the origin than $z_s$.
It turns out that they cannot separate along the real axis either (case 
$\Delta_3<0$), because the integrals of Eq. \ref{invcsatrK}
would have a negative sign. The only remaining possibility is thus
that for small enough $p_1$, there remains a double real solution $z_d$ 
corresponding to the condition 
 of the vanishing determinant
\begin{eqnarray}
\Delta_3  = 0
\label{Delta3zero}
\end{eqnarray}
Then the simple real solution $z_s$ and the double real solution $z_d $ read
\begin{eqnarray}
 z_s && = \frac{3 Q}{P}-\frac{b}{3a}
\nonumber \\
 z_d && = - \frac{3 Q}{2 P}-\frac{b}{3a}
\label{zszd}
\end{eqnarray}
and $R(z)$ can be rewritten as
\begin{eqnarray}
R(z)   = \frac{p_1}{p_2} (z_s-z) (z-z_d)^2 
\label{rzd3zero}
\end{eqnarray}
i.e. there exists a single cut $[z_s,+\infty[$ on the real axis.
After the deformation of the contour in the complex plane, the coefficients
of Eq. \ref{invcsatrK} can be written as the real integral
\begin{eqnarray}
c_{\infty}(k) && = \oint_{C_0} \frac{d z }{2i \pi z^{k+1} } {\hat c}_{\infty}( z)
=  \frac{\sqrt{p_2 p_1 {\hat c}_{\infty} (0)} }{\pi}
\int_{z_s}^{+\infty} dx \frac{ \sqrt{x-z_s} (x-z_d) }{x^k (p_2+p_1 x^2)^2}
\label{invcsatrKcut}
\end{eqnarray}
which is positive for $z_s \geq z_d$, which should be possible for small enough $p_1$
since one starts from the values of Eq. \ref{rzp1zero}.
In particular, the asymptotic behavior for large $k$ reads
\begin{eqnarray}
c_{\infty}(k) && \oppropto_{k \to +\infty} 
  \frac{\sqrt{p_2 p_1 {\hat c}_{\infty} (0)} \Gamma \left(\frac{3}{2} \right) (z_s-z_d)}
{\pi (p_2+p_1 z_s^2)^2 z_s^{k-3/2} k^{3/2}}
\label{cinftyklarge}
\end{eqnarray}

Let us now study more precisely the condition $\Delta_3=0$ of Eq. \ref{Delta3zero} :
this means that $ {\hat c}_{\infty} (0) $ has to be one root of the cubic polynomial $N_3$
of Eq. \ref{Delta3}.
Let us start from the simple case $p_1=1-p_2=0 $
\begin{eqnarray}
N_3^{(p_1=0)}   = 1 - {\hat c}_{\infty} (0) 
\label{n3special}
\end{eqnarray}
that vanishes for ${\hat c}_{\infty} (0)=1 $, which is indeed the correct solution 
(Eq \ref{solup2eq1}). 
In the region $0 \leq p_1 < \frac{1}{28}$, the discriminant $\delta_3$ of Eq. \ref{Delta3n3}
remains negative, so that the three real roots of Eq. \ref{n3t}
can be written using the integers  $m=0,1,2$ as
\begin{eqnarray}
t_m= 2 \sqrt{ \frac{-p}{3} } \cos \left[ \frac{\theta+2 \pi m}{3} \right]
\label{3reelles}
\end{eqnarray}
where the angle $0<\theta<\pi$ satisfies
\begin{eqnarray}
\cos \theta = \frac{- q }{2} \sqrt{\frac{27}{-p^3}} 
= - (63-280 p_1+3584p_1^2-4096 p_1^3)
 \sqrt{\frac{ p_1   }{ p_2 (3 + 64 p_1 (1-4 p_1) )^3}}
\label{costheta}
\end{eqnarray}
In the limit  $p_1 =1-p_2 \to 0$, we have at first order in $p_1$
\begin{eqnarray}
p && =- \frac{1}{64 p_1} - \frac{67}{92} +O(p_1)
\nonumber \\
q && = \frac{7}{768 p_1} - \frac{77}{3456} +O(p_1)
\nonumber \\
\cos \theta && = - 7 \sqrt{3 p_1}   +O(p_1^{3/2})
\nonumber \\
\theta  && =  \frac{\pi}{2}+7 \sqrt{3 p_1}   +O(p_1^{3/2})
\label{pqdvp1}
\end{eqnarray}
so that the only root that remains finite as $p_1 \to 0$ corresponds to the value $m=2$ 
in Eq. \ref{3reelles}
\begin{eqnarray}
t_2= \frac{7}{12}- \frac{7}{4} p_1 +O(p_1^2)
\label{t2dv}
\end{eqnarray}
whereas the two other solutions flow towards $(\pm \infty)$
\begin{eqnarray}
t_0 &&  =  \frac{1}{8 \sqrt{p_1}}- \frac{7}{24} +O(p_1^{1/2})
\nonumber \\
t_1 &&  = - \frac{1}{8 \sqrt{p_1}}- \frac{7}{24} +O(p_1^{1/2})
\label{t01dv}
\end{eqnarray}
The 'physical' solution for ${\hat c}_{\infty}(0) $ (Eq \ref{ct}) has thus for expansion
\begin{eqnarray}
{\hat c}_{\infty}(0)=C_2=t_2 + \frac{ (8p_2-3)}{12 p_2} = 1- 2 p_1 +  +O(p_1^2)
\label{cdv}
\end{eqnarray}
The two other unphysical roots of $\Delta_3$ read
\begin{eqnarray}
C_0 &&  =  t_0 + \frac{ (8p_2-3)}{12 p_2} = \frac{1}{8 \sqrt{p_1}}+ \frac{1}{8} +O(p_1^{1/2})
\nonumber \\
C_1 &&  = t_1 + \frac{ (8p_2-3)}{12 p_2}= - \frac{1}{8 \sqrt{p_1}}+ \frac{1}{8} +O(p_1^{1/2})
\label{c01dv}
\end{eqnarray}
By continuity in $p_1$, we thus expect that ${\hat c}_{\infty}(0)=C_2$ 
in the whole region $0 \leq p_1 < \frac{1}{28}$ where the three real roots of $\Delta_3$
remain distinct. At the border $p_1 = \frac{1}{28} $, the discriminant $\delta_3$ vanishes :
there exists a double root ${\hat c}_{\infty} (0)=49/54$ (coalescence of the two solutions $C_2$ et $C_0$)
and a simple root $C_1=-16/27$ that is negative.
For $p_1 > \frac{1}{28} $, the discriminant $\delta_3$ becomes positive, 
i.e. the double root $C_0=C_2=49/54$ become a pair of complex conjugated roots
(this cannot correspond to any physical  ${\hat c}_{\infty} (0)$ which has to be real),
and the simple root $C_1 $ remains real (but this cannot correspond to any physical  ${\hat c}_{\infty} (0)$ which has to be positive).

So we have found that a stationary solution ${\hat c}_{\infty} (k)$ exists in the region 
$0 \leq p_1 \leq \frac{1}{28}$. Let us consider more precisely the solution
at the frontier $p_1 = \frac{1}{28} $ where we have found ${\hat c}_{\infty} (0)=49/54$ :
it turns out that the parameters of Eq. \ref{eqxPQ} both vanish at this point
 $P=0=Q$, i.e. there exists a single triple root to $R(z)$
 (merging of the simple solution $z_s$ and of the double solution $z_d$
existing in the region $0 \leq p_1 < \frac{1}{28}$)
\begin{eqnarray}
z_s=z_d=3
\label{ztriplespecial}
\end{eqnarray}
Then Eq \ref{invcsatrKcut} simplifies into
\begin{eqnarray}
c_{\infty}(k) && 
=  \frac{ 4.7^2 }{\pi  \sqrt{2} }
\int_{3}^{+\infty} dx \frac{ (x-3)^{3/2}  }{x^k (27+ x^2)^2}
\label{invcsatrKcutspecial}
\end{eqnarray}
It is easy to check that it satisfies the value at $k=0$
\begin{eqnarray}
c_{\infty}(k=0) && = \frac{ 4.7^2 }{\pi  \sqrt{2} }
\int_{3}^{+\infty} dx \frac{ (x-3)^{3/2}  }{ (27+ x^2)^2}
=  \frac{49}{54}
\end{eqnarray}
and the normalization condition 
\begin{eqnarray}
1  = \sum_{k=0}^{+\infty} c_{\infty}(k)
=  \frac{ 4.7^2 }{\pi  \sqrt{2} }
\int_{3}^{+\infty} dx \frac{ x (x-3)^{3/2}  }{(x-1) (27+ x^2)^2}
\end{eqnarray}

In particular, the asymptotic behavior for large $k$ of Eq. \ref{invcsatrKcutspecial} reads
\begin{eqnarray}
c_{\infty}(k) && \oppropto_{k \to +\infty} 
  \frac{\sqrt{p_2 p_1 {\hat c}_{\infty} (0)} \Gamma \left(\frac{5}{2} \right) }{\pi (p_2+p_1 z_s^2)^2
  z_s^{k-5/2} k^{5/2}}
\label{cinftyklargecriti}
\end{eqnarray}
in contrast to Eq. \ref{cinftyklarge}.

It is thus interesting to mention now the singularities that appear for  
\begin{eqnarray}
p_1= \frac{1}{28} - \epsilon \ \ {\rm with } \ \ \epsilon>0
\label{epsneg}
\end{eqnarray}
We find
\begin{eqnarray}
p && = - \frac{3}{4}  - \frac{1078}{81} \epsilon +...
\nonumber \\
q && =  \frac{1}{4}  + \frac{539}{81} \epsilon+...
\nonumber \\
\delta_3 && = - \frac{4519603984 }{14348907} \epsilon^3 +...
\nonumber \\
\cos \theta && = -1 +\frac{36156831872}{14348907} \epsilon^3
\nonumber \\
\theta && = \pi - \frac{268912 }{2187 \sqrt{3}} \epsilon^{3/2}
\label{Dvmoinseps}
\end{eqnarray}
and
\begin{eqnarray}
c_{\infty}(0) = C_2 && = \frac{49}{54} + \frac{3430}{729} \epsilon - \frac{134456 }{6561} \epsilon^{3/2} +..
\nonumber \\
C_1  && = - \frac{16}{27} - \frac{6272}{729}  \epsilon+..
\nonumber \\
C_0 = && =\frac{49}{54} + \frac{3430}{729} \epsilon + \frac{134456 }{6561} \epsilon^{3/2} +..
\label{dvcoepsneg}
\end{eqnarray}
Finally, the parameters $P$ and $Q$ (Eq \ref{eqxPQ}) 
and the roots $(z_s,z_d)$ (Eq \ref{rzd3zero}) of the cubic equation in $z$ (Eq. \ref{Rz}),
read
\begin{eqnarray}
P && = - \frac{784}{3} \epsilon+...
\nonumber \\
Q  && = - \frac{43904}{27} \epsilon^{3/2}+...
\nonumber \\
z_s && = 3+ \frac{56}{3}  \epsilon^{1/2}
\nonumber \\
z_d && = 3- \frac{28}{3}  \epsilon^{1/2}
\end{eqnarray}
In particular, the difference $(z_s-z_d)$ between the simple and the double root 
presents a square-root singularity in $\sqrt{\epsilon}$.
Let us now discuss what happens on the other side of $p_1=1/28$.

\subsection { Region  $p_1=  \frac{1}{28} +\epsilon $ with $\epsilon>0$ }

\label{quasistatio}

\begin{figure}[htbp]
 \includegraphics[height=6cm]{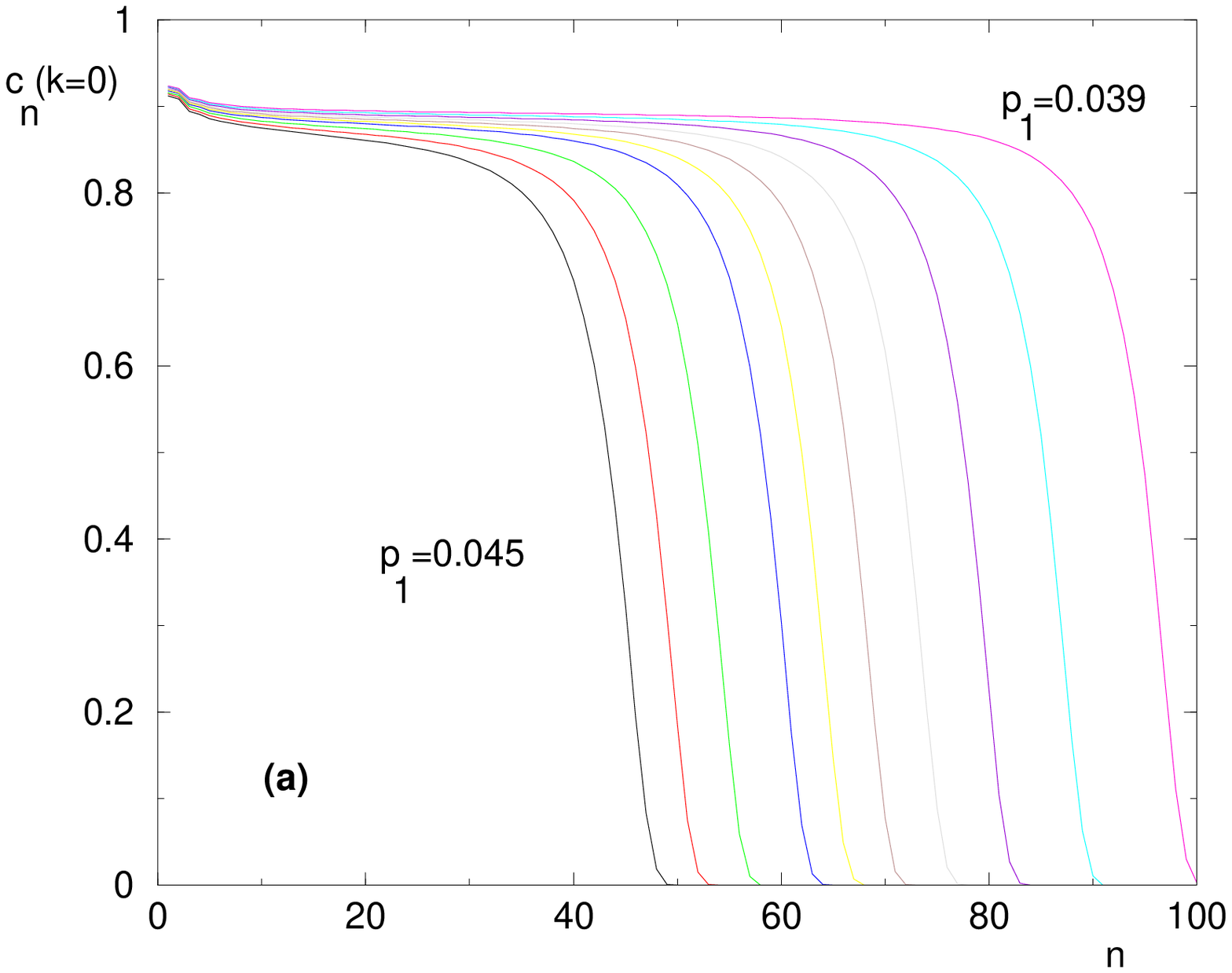}
\hspace{2cm}
\includegraphics[height=6cm]{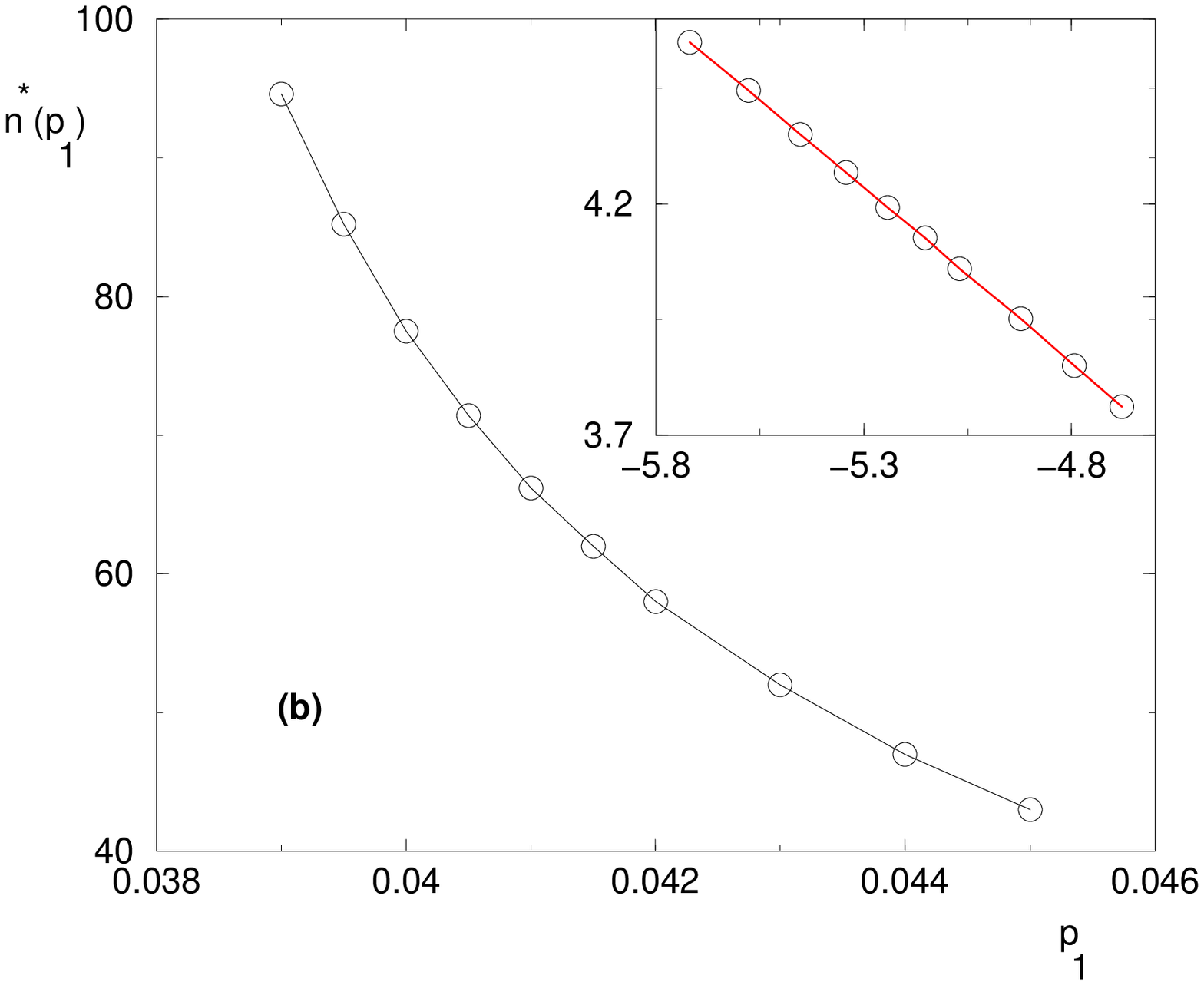}
\caption{ Region $p_1= \frac{1}{28} +\epsilon $ with $\epsilon>0$
(a) $c_n(k=0)$ as a function of $n$ for various values of $p_1$ : there is a rapid convergence towards a finite value, and then a decay towards zero at some $n^*(p_1)$ 
(b) Divergence of the scale $n^*(p_1)$ as $p_1 \to 1/28 = 0.0357$:   the inset displays the same data in the log variables $\ln n^*(p_1)$ as a function of $\ln \left(p_1-\frac{1}{28} \right)$. We measure a slope of order 
$\nu_+ \simeq 0.75 $ (see Eq. \ref{dvnetoile}). }
\label{figquasi}
\end{figure}

For $p_1=  \frac{1}{28} +\epsilon $ with $\epsilon>0$, as shown on Fig. \ref{figquasi}
we find numerically from Eq. \ref{recgenecK}
that $c_n(k)$ rapidly converges towards a quasi-stationary metastable distribution
$c_*(k)$ that disappears after a number of generation $n_*(\epsilon)$ that diverges as
$\epsilon \to 0$ as
\begin{eqnarray}
n_*(p_1) \oppropto_{p_1 \to \left( \frac{1}{28} \right)^+} \frac{1}{\left(p_1-\frac{1}{28} \right)^{\nu_+}} \ \ 
{\rm with }  \ \ \nu_+ \simeq 0.75
\label{dvnetoile}
\end{eqnarray}
This diverging length represents the radial distance near the boundary where
 the variable $k$ remains finite, i.e. where
the quantum ferromagnetic clusters remain finite. But then after $n_*$, the variable $k$ flows towards infinity,
i.e. an extensive ferromagnetic cluster is formed. For a tree of $n$ generations, the renormalized transverse field of this central cluster will scale as
\begin{eqnarray}
\ln h_R(n) \propto - k \propto - K^{n-n_*}
\label{scalingketoile}
\end{eqnarray}
This should be compared to the scaling $\ln h_R(L) \propto - \left( \frac{L}{\xi_h}\right)^d$ in finite dimension $d$. 
The corresponding renormalized external ferromagnetic coupling will scale as
\begin{eqnarray}
J^{Rext}_n \simeq K^{n-n_*} J^{Rext}_{n_*}
\label{scalingjextetoile}
\end{eqnarray}
where $J^{Rext}_{n_*} $ corresponding to the first $n_*$ generations
has the behavior of a modified Directed Polymer partition function as we explain in the following section.
This should be compared to the scaling $ J^{ext}_L \propto  \sigma L^{d-1}$ in finite dimension $d>1$ where there exists an underlying
classical transition (see \cite{melin_tree} and references therein for the properties of the classical Ising model on the Cayley tree).

\subsection{ Modified Directed Polymer model for $J^{ext}$ when there exists a finite $c_{\infty}(k)$ }

In the region $J<h_{min}$ where only sites are decimated and where 
the asymptotic distribution $c_{\infty}(k) $ reduces to the delta function
$c_{\infty}(k)=\delta_{k,0} $, we have seen that the renormalization of the external coupling $J_{ext}$
corresponds quantitatively to a Directed Polymer model described in section  \ref{sec_decisites}.
In the region $h_{min}<J<h_{max}$ where the asymptotic distribution $c_{\infty}(k) $
is not a delta function anymore but remains a finite distribution with no rescaling in $k$,
the renormalization of the external coupling $J_{ext}$
still corresponds to a Directed Polymer model, but with a slightly different disorder
with respect to the case described in section  \ref{sec_decisites} :
the effective random energies of Eq. \ref{epssite}
that were both positive for the binary case in the region $J<h_{min}$,
with $\epsilon_1= \ln \frac{h_1}{J}$ with probability $p_1$ and 
$\epsilon_2= \ln \frac{h_2}{J}$ with probability $p_2$, are replaced 
in the region $h_{min}<J<h_{max}$  when there exists a finite $c_{\infty}(k)$ 
\begin{eqnarray}
\epsilon_1' && = 0 \ \ {\rm with \ \ probability} \ \ p_1'=1-p_2 c_{\infty}(k=0) \nonumber \\
\epsilon_2' && = \ln \frac{h_2}{J}= \ln \sqrt{\frac{h_2}{h_1}}  \ \ {\rm with \ \ 
probability} \ \ p_2'=p_2 c_{\infty}(k=0)
\label{epssitemodified}
\end{eqnarray}
The factor $ c_{\infty}(k=0)$ enters because the non-trivial contribution to the renormalization of $J_{ext}$ occurs only if the corresponding site gets decimated, i.e. one needs both conditions $h_i=h_2$ and $k=0$ 
(see the renormalization rules of section \ref{sec_rgbranch}).

The function of Eq. \ref{fbeta} has thus to be modified into
 \begin{eqnarray}
 f_{mod}(\beta)  \equiv - 
\frac{1}{\beta} \ln \left( K \left[ (1-p_2 c_{\infty}(k=0) )
+p_2 c_{\infty}(k=0) \left( \frac{h_1}{h_2} \right)^{\frac{\beta}{2}}  \right] \right)
\label{fbetamodified}
  \end{eqnarray}

Again for $1> \beta_c $, the quantum model can then only be
in its disordered phase, whereas for $1< \beta_c$, 
the criticality condition $ f_{mod}(\beta=1)$ for the quantum model of Eq. \ref{jcquantdeloc}
becomes
 \begin{eqnarray}
1 = K \left[ (1-p_2 c_{\infty}(k=0) )
+p_2 c_{\infty}(k=0) \left( \frac{h_1}{h_2} \right)^{\frac{1}{2}}  \right]
\label{jcquantdelocmodified}
  \end{eqnarray}
and can only occur within the delocalized phase of the Directed Polymer.

\subsection{ Summary }

The analysis of the region $h_{min}<J<h_{max}$
can be summarized as follows :

(i) either the quantum transition occurs in the region where
there exists a finite asymptotic distribution $c_{\infty}(k)$
describing {\it finite quantum ferromagnetic clusters} :
then the transition is analog to the transition that can occur in the region $J<h_{min}$
(see section \ref{sec_decisites})
and is determined by the modified Directed Polymer model described just above.
The critical behaviors are then governed again by Eqs \ref{jldpdelocdes} and Eq \ref{jldpdelocxityp}
with the critical exponent $\nu_{typ}=1$.

(ii) or the quantum model remains disordered as long as
there exists a finite asymptotic distribution $c_{\infty}(k)$, i.e. up to the critical point $p_1=1/28$ 
described above. It
becomes ordered only
 when there appears an {\it infinite quantum ferromagnetic cluster} 
for $p_1=1/28+\epsilon$.
Then the critical behaviors are completely different from (i) and are determined by
the scaling of section \ref{quasistatio} : the finite-size renormalized transverse field and 
renormalized external coupling scale as Eq. \ref{scalingketoile}
and \ref{scalingjextetoile} where the diverging length $n_*$ of Eq. \ref{dvnetoile}
involves the critical exponent $\nu_+\simeq 0.75$.

\section{ Conclusion }

In this paper, we have considered the  Random Transverse Field Ising model on a Cayley tree.
To avoid the difficulties of the usual Strong Disorder RG that destroys the tree structure,
we have introduced a modified procedure called 'Boundary Strong Disorder Renormalization'
that preserves the tree structure, so that one can write simple recursions with respect to the number of 
generations. We have first checked that this modified procedure allows to recover exactly most of the 
critical exponents for the one-dimensional chain. Then we have studied the RG equations
for the Cayley tree with a uniform ferromagnetic coupling $J$ and random transverse fields within the support $[h_{min},h_{max}]$. We have found the following picture :

(i) for $J>h_{max}$, only bonds are decimated, so that the whole tree is a quantum ferromagnetic cluster

(ii) for $J<h_{min}$, only sites are decimated, so that no quantum ferromagnetic cluster is formed, 
and the ferromagnetic coupling to the boundary corresponds quantitatively to the partition function of a Directed Polymer model.
This relation with the Directed Polymer has been already obtained for the Cayley tree via some approximations
within the Quantum Cavity Approach  \cite{ioffe,feigelman,dimitrova}, and for arbitrary networks via 
simple perturbation deep in the disordered phase \cite{us_transverseDP}. However here we also consider the
possibility of bond-decimations to build quantum ferromagnetic clusters which may become important near the transition.

 (iii) for $h_{min}<J<h_{max}$, both sites and bonds can be decimated. When the quantum ferromagnetic clusters remain finite, the physics is similar to (ii), with a quantitative mapping to a modified Directed Polymer model; otherwise an extensive quantum ferromagnetic cluster appears.
 
We have found that the quantum transition can be of two types :

(a) either the quantum transition takes place in the region where quantum ferromagnetic clusters do not exist or remain finite.
Then the ferromagnetic coupling to the boundary behaves in the disordered phase as 
$J^{ext}_n \sim e^{- n/\xi_{typ}}$
where the correlation length $\xi_{typ}$ diverges with the typical correlation 
length exponent $\nu_{typ}=1$.

(b) or the quantum transition takes place at the point 
where an extensive quantum ferromagnetic cluster appears at the center of the tree, at a radial distance $n_*$
from the boundary that diverges with the correlation exponent $\nu_+\simeq 0.75$. In the ordered phase, 
the finite-size renormalized transverse field and 
renormalized external coupling scale as
$\ln h_R(n) \propto  - K^{n-n_*}$ and $J^{Rext}_n \simeq K^{n-n_*} J^{Rext}_{n_*} $.

In a companion paper \cite{boundary2d}, we describe how the idea of Boundary Strong Disorder RG can be adapted in dimension $d=2$ and we present the corresponding numerical results.

\appendix

\section{ Reminder on Strong Disorder RG rules on arbitrary lattices }

\label{app_full}

In this section, we recall the standard Strong Disorder Renormalization for 
 the Random Transverse Field Ising Model of Eq. \ref{hdes}.


For the model of Eq. \ref{hdes},
the Strong Disorder RG rules are formulated on arbitrary lattices as follows \cite{fisherreview,motrunich} :

(0) Find the maximal value among the transverse fields $h_i$
and the ferromagnetic couplings $J_{jk}$
\begin{eqnarray}
\Omega= {\rm max } \left[h_i,J_{jk} \right]
\label{omega}
\end{eqnarray}

i) If $\Omega= h_{i}$, then the site $i$ is decimated and disappears,
while all couples $(j,k)$ of neighbors of $i$ are now linked via
the renormalized ferromagnetic coupling
\begin{eqnarray}
J_{jk}^{new} = J_{jk} + \frac{ J_{ji} J_{ik} }{h_{i}}
\label{jjknew}
\end{eqnarray}

ii) If $\Omega= J_{ij}$, then the site $j$ is merged with the site $i$.
The new renormalized site $i$ has a reduced renormalized transverse field
\begin{eqnarray}
h_{i}^{new}= h_i r_i \ \ {\rm with } \ \ r_i= \frac{h_j}{ J_{ij}  }
\label{hinew}
\end{eqnarray}
and a bigger magnetic moment
\begin{eqnarray}
 \mu_{i}^{new}=\mu_{i}+\mu_j
\label{minew}
\end{eqnarray}
This renormalized cluster is connected to other sites via the renormalized couplings
\begin{eqnarray}
J_{ik}^{new} =  J_{ik}+J_{jk}
\label{jiknew}
\end{eqnarray}

(iii) return to (0).

\end{document}